\documentclass[11pt,a4paper]{article}
\usepackage{amsfonts,amscd,amsmath,amssymb,amsthm,times}
\newtheorem{lem}{Lemma}[section]
\newtheorem{theo}{Theorem}[section]

\newtheorem{cor}{Corollary}[section]
\newtheorem{ex}{Example}[section]
\newtheorem{re}{Remark}[section]

\DeclareMathOperator{\Ret}{Re}
\DeclareMathOperator{\Imt}{Im}
\DeclareMathOperator{\tr}{Tr}
\DeclareMathOperator{\Ran}{Ran}
\DeclareMathOperator{\Ker}{Ker}
\DeclareMathOperator{\Rank}{Rank}

\newcommand{\R}{\mathbb{R}}
\newcommand{\C}{\mathbb{C}}
\newcommand{\Z}{\mathbb{Z}}
\newcommand{\N}{\mathbb{N}}

\newcommand{\1}{\mathbb{I}}

\newcommand{\cD}{{\cal D}}
\newcommand{\cH}{{\cal H}}
\newcommand{\cM}{{\cal M}}
\newcommand{\cU}{{\cal U}}
\newcommand{\cE}{{\cal E}}
\newcommand{\cI}{{\cal I}}
\newcommand{\cV}{{\cal V}}
\newcommand{\cW}{{\cal W}}
\newcommand{\cG}{{\cal G}}
\newcommand{\cK}{{\cal K}}
\newcommand{\U}{\mathsf{U}}
\newcommand{\SL}{\mathsf{SL}}

\title{Kirchhoff's Rule for Quantum Wires}
\author{V. Kostrykin\thanks{e-mail: kostrykin@t-online.de, kostrykin@ilt.fhg.de}\\
Lehrstuhl f\"ur Lasertechnik\\ Rheinisch
- Westf\"alische Technische Hochschule Aachen\\
Steinbachstr. 15, D-52074 Aachen, Germany
\and
and\\
 R. Schrader\thanks{e-mail: schrader@physik.fu-berlin.de, Supported in part by
DFG SFB 288 ``Differentialgeometrie und Quantenphysik''}\\ Institut f\"{u}r
Theoretische Physik\\ Freie Universit\"{a}t Berlin, Arnimallee 14\\ D-14195 Berlin,
Germany}

\begin{document}
\maketitle

\centerline{\large Dedicated to S.P. Novikov on the occasion of his 60-th birthday}

\begin{abstract}
In this article we formulate and discuss one particle quantum scattering theory
on an arbitrary finite graph with $n$ open ends and where we define the
Hamiltonian to be (minus) the Laplace operator with general boundary conditions
at the vertices. This results in a scattering theory with $n$ channels. The
corresponding on-shell S-matrix formed by the reflection and transmission
amplitudes for incoming plane waves of energy $E>0$ is explicitly given in
terms of the boundary conditions and the lengths of the internal lines.  It is
shown to be unitary, which may be viewed as the quantum version of Kirchhoff's
law. We exhibit covariance and symmetry properties. It is symmetric if the
boundary conditions are real. Also there is a duality transformation on the set
of boundary conditions and the lengths of the internal lines such that the low
energy behaviour of one theory gives the high energy behaviour of the
transformed theory. Finally we provide a composition rule by which the on-shell
S-matrix of a graph is factorizable in terms of the S-matrices of its
subgraphs. All proofs only use known facts from the theory of self-adjoint
extensions, standard linear algebra, complex function theory and elementary
arguments from the theory of Hermitian symplectic forms.
\end{abstract}

\footnotetext[1]{Published in
\textit{J. Phys. A: Math. Gen.} \textbf{32} (1999), 595-630.}

\newpage

\section{Introduction}

At present mesocopic quasi-one-dimensional structures like quantum \cite{qw,Serena},
atomic \cite{Joachim} and molecular \cite{mw} wires have become the subject of  intensive
experimental and theoretical studies. This kind of electronics is still far from being
commercially useful. However, the enormous progress that has been made in the past years
suggests that it will not be too long before the first molecule-sized electronic components
become a reality (see e.g. \cite{Sols1,Sols2,Kouwenhoven}).

According to already traditional physical terminology a quantum wire is a graph-like
structure on a surface of a semiconductor, which confines an electron to potential grooves
of width of about a few nanometers. An accurate theory for these nanostructures must
include confinement, coupling between closely spaced wires, rough boundaries, impurities,
\textsl{etc}. The simplest model describing the conduction in quantum wires is a
Hamiltonian on a planar graph. A similar model can be applied to molecular wire
-- a quasi-one-dimensional molecule that can transport charge carriers
(electrons or holes) between its ends \cite{Ruedenberg}. Atomic wires, i.e.
lines of metal atoms on the surface of a semiconductor provide another example
of such quasi-one-dimensional structures. Also Hamiltonians on planar graphs
arise naturally in the modelling of high-temperature granular superconductors
\cite{Gennes,Alex,Jeffery}.

Although such models were proposed long ago (see e.g.
\cite{Griffith1,Griffith2}), probably it was Pavlov and Gerasimenko
\cite{Gerasimenko,Gerasimenko2} who initiated a rigorous mathematical analysis
of such models, which later acquired the name of quantum wires. A more general
approach to the problem of the corresponding mathematical structure was
formulated in \cite{Livsic} several decades earlier. Here we do not intend to
give a complete overview of the whole subject. We only mention some related
studies. In \cite{Avron:Sadun1} and \cite{Avron:Sadun2} networks with leads
were used to study adiabatic transport and Chern numbers. Two particle
scattering theory on graphs was studied in \cite{Melnikov:Pavlov}. Quantum
waveguides
\cite{Schult,Avishai,Berggren:Ji,Duclos:Exner,Simon,Carini1,Carini2}, where the
influence of confining potentials walls is modeled by the Dirichlet boundary
condition, give a more realistic description of quasi-one dimensional
conductors. The wave function is allowed to have several mutually interacting
transversal modes. In real quantum wires the number of these modes can be
rather large (up to $10^2$
-- $10^3$). For another more realistic model of a two dimensional quantum wire
see \cite{Wu:Sprung}.

In this article we consider idealized quantum wires, where the configuration
space is a graph, i.e. a strictly one-dimensional object and the Hamiltonian is
minus the Laplacian with arbitrary boundary conditions at the vertices of the
graph and which makes it a self-adjoint operator. The graph need not to be
planar and may be bent when realized as a subset of the 3-dimensional Euclidean
space $\R^{3}$. By now many explicit examples have been considered (see e.g.
\cite{Gerasimenko,Avron:Raveh:Zur,Exner:Seba:88,Exner:Seba:Stovicek,Exner89,Adamyan,
Avishai:Luck,Avron:Exner:Last,Carreau,Gratus,Jonckheere}) including also the
Dirac operator with suitable boundary conditions \cite{Bulla:Trenckler}. Our
approach gives a systematic discussion and covers in particular all these cases
for the Laplace operator. In this article we will, however, not be concerned
with the question, which of these boundary conditions could be physically
reasonable. The physical relevance of different boundary conditions is
discussed e.g. in \cite{Exner89,Exner96a}.

The scattering theory for these operators exhibit a very rich structure (see e.g.
\cite{Tekman:Bagwell,Porod:Shao:Lent,Kottos:Smilansky}) and by Landauer's theory
\cite{Landauer} provides the background for understanding conductivity in mesoscopic
systems. The on-shell S-matrix at energy $E$ is an $n\times n$ matrix if the graph has $n$
open ends, which we will show to be given in closed matrix form in terms of the boundary
conditions and the lengths of the internal lines of the graph. We exhibit covariance and
invariance properties and show in particular that the on-shell S-matrix is symmetric for
all energies if the boundary conditions are real in a sense which we will make precise.
The main result of this article is that this on-shell S-matrix is unitary, continuous in
the energy and even real analytic except for an at most denumerable set of energies
without finite accumulation points. This set is given in terms of the boundary conditions
and the lengths of the internal lines. This result may be viewed as the quantum version of
Kirchhoff's rule. For explicit examples this has been known (see e.g.
\cite{Exner94,Exner96a}), but again our approach provides a unified treatment. Physically
this unitarity is to be expected since there is a local Kirchhoff rule. In fact, the
boundary conditions imply that the quantum probability currents of the components of any
wave packet associated to the different lines entering any vertex add up to zero. Our
discussion of the boundary conditions will be based on Green's theorem and will just
reflect this observation. We will actually give three different proofs, each of which will
be of interest in its own right.

Finally there is a general duality transformation on the boundary conditions (turning
Dirichlet into Neumann boundary conditions and vice versa) which combined with an energy
dependent scale transformation on the lengths of the internal lines relates the high
energy behaviour of one theory to the low energy behaviour of the transformed theory.

This article is organized as follows. In Section 2 we discuss the simple case
with one vertex only but with an arbitrary number of open lines ending there.
This will allow us to present the main elements of our strategy, which is the
general theory of self-adjoint extensions of symmetric operators and its
relation to boundary conditions in the context of Laplace operators. This
discussion uses some elementary facts about Hermitian symplectic forms.
Although some results will be proven again for the general set-up in Section 3,
for pedagogical reasons and because they are easier and more transparent in
this simple case, we will also give proofs. In Section 3 we discuss the general
case with the techniques and mostly with proofs, which extend those of Section
2. We start with a general algebraic formulation of boundary conditions
involving a finite set of half lines and finite intervals but without any
reference to local boundary conditions on a particular graph. At the end we
show that any of these boundary conditions may be viewed as local boundary
conditions on a suitable (maximal) graph.

The connection between the theory of self-adjoint extensions of symmetric
operators and Hermitian symplectic forms was brought to the attention of one of
the authors (R.S.) by G. Segal in 1987. In a recent paper \cite{Novikov} S.P.
Novikov stated that he learned this from I.M. Gelfand back in 1971.
Unfortunately we were unable to trace back the precise history of this
connection. It seems that it was made by several researchers at different times
(see e.g. \cite{Kochubei1,Kochubei,Pavlov2,Kurasov:Boman}) but still was not
analyzed systematically. In Sections 2 and 3 and in Appendix A we will try to
fill this gap.

In Section 4 we consider the question what happens if one decomposes a graph
into two or more components by cutting some of its internal lines and replacing
them by semi-infinite lines. One would like to compare the on-shell S-matrices
obtained in this way with the original one. If the graph has two open ends and
and its subgraphs are connected by exactly one line, the Aktosun factorization
formula \cite{Aktosun} (see also
\cite{Redheffer,Redheffer2,Kowal,Rozman,Aktosun:Klaus:Mee,Bianchi,Bianchi2,Exner:Tater,Jung,KS1})
for potential scattering on the line easily carries over to this case. Such
rules are reminiscent of the Cutkosky cutting rules \cite{Cutkosky} for
one-particle reducible Feynman diagrams. Also such relations are well known in
network theories and then the composition law for the S-matrices figures under
the name star product \cite[pp. 285-286]{Redheffer} (we would like to thank M.
Baake (T\"ubingen) for pointing out this reference), \cite{Redheffer2}. If the
cutting involves more than one line, the situation becomes more complicated and
leads to interesting phenomena related to the semiclassical Gutzwiller formula
and the Selberg trace formula \cite{Kottos:Smilansky} (see also
\cite{Avron:Sadun1,Avron:Sadun2,Kowal}). We provide a general composition rule
for unitary matrices, which we will call a generalized star product and by
which the on-shell S-matrix of an arbitrary graph (with local boundary
conditions) can be factorized in terms of the S-matrices of its subgraphs. We
expect that this general, highly nonlinear composition rule could also be of
relevance in other contexts. Note that there is some similarity between our
results and the recursive approach of \cite{Shi:Gu}. Section 5 contains a
summary and an outlook for possible applications and further investigations.

When this article was already submitted for publication we have learned about
the work \cite{Novikov} (some results of which were announced in a short note
\cite{Novikov2}) by S.P. Novikov, where a program similar to ours is carryied
out for discrete (combinatorial) Laplacians on tree graphs.

\section{The Quantum Wire with a Single Vertex}

In this section we will consider a quantum wire with $n$ open ends and joined at a single
vertex. This toy model will already exhibit most of the essential features of the general
case and is also of interest in its own right. In particular the general strategy and the main
techniques of our approach will be formulated in this section. Let the Hilbert space be
given as
\begin{equation*}
\cH=\oplus^{n}_{i=1}\cH_{i}=\oplus^{n}_{i=1}L^{2}([0,\infty)).
\end{equation*}
Elements $\psi\in\cH$ will be written as $(\psi_{1},\psi_{2},...,\psi_{n})$ and we will
call $\psi_{j}$ the component of $\psi$ in channel $j$. The scalar product in $\cH$ is
\begin{equation*}
 \langle\phi,\psi\rangle=\sum^{n}_{i=1}\langle\phi_{i},\psi_{i}\rangle_{\cH_i}
\end{equation*}
with the standard scalar product on $L^{2}([0,\infty))$ on the right hand side. We
consider the symmetric operator $\Delta^0$ on $\cH$, such that
\begin{displaymath}
\Delta^0\psi=\left(\frac{d^2\psi_1}{dx^2},\ldots,\frac{d^2\psi_n}{dx^2}\right)
\end{displaymath}
with domain of definition $\cD (\Delta^{0})$ being the set of all $\psi$ where
each $\psi_i$, $1\leq i\leq n$ together with their first and second generalized
derivatives belong to $L^2(0,\infty)$ (i.e. $\psi_i\in W^{2,2}(0,\infty)$, a
Sobolev space) and which vanish at $x=0$ together with their first derivatives.
It is clear that $\Delta^{0}$ has defect indices $(n,n)$, such that the set of
all self-adjoint extensions can be parametrized (in a noncanonical way) by the
unitary group $\U(n)$, which has real dimension $n^{2}$ (see e.g.
\cite{Achieser:Glasmann,RS}).

There is an alternative and equivalent description of all self-adjoint
extensions in terms of symplectic theory and which goes as follows. Let
$\cD\subset\cH$ be the set of all $\psi$ such that each $\psi_{i}$, $1\le i\le
n$ belongs to $W^{2,2}(0,\infty)$
- and we will then say $\psi\in W^{2,2}$.
 On $\cD$ consider the following skew-Hermitian quadratic form
given as
\begin{equation*}
 \Omega(\phi, \psi)= \langle\Delta\phi,\psi\rangle-\langle\phi,\Delta\psi\rangle
=-\overline{\Omega(\psi,\phi)}
\end{equation*}
with the Laplace operator $\Delta=d^2/dx^2$ considered as a differential
expression. Obviously $\Omega$ vanishes identically on $\cD(\Delta^{0})$. Any
self-adjoint extension of $\Delta^{0}$ is now given in terms of a maximal
isotropic subspace of $\cD$, i.e. a maximal (linear) subspace on which $\Omega$
vanishes identically. This notion corresponds to that of Lagrangean subspaces
in the context of Euclidean symplectic forms (see e.g. \cite{Arnold}). To find
these maximal isotropic subspaces we perform an integration by parts (Green's
theorem) and obtain with $^{\prime}$ denoting the derivative and $\;\bar{}\;$
complex conjugation
\begin{equation*}
 \Omega(\phi,\psi)=\sum^{n}_{i=1}\left(\bar{\phi}_{i}(0)\psi^{\prime}_{i}(0)
                               -\bar{\phi}^{\prime}_{i}(0)\psi_{i}(0)\right).
\end{equation*}
We rewrite this in the following form. Let $ [\:]: \cD\rightarrow \C^{2n}$
be the surjective linear map which associates to $\psi$ and $\psi^{\prime}$
their boundary values at the origin:
\begin{equation*}
[\psi]=(\psi_{1}(0),...\psi_{n}(0),
       \psi^{\prime}_{1}(0),...\psi^{\prime}_{n}(0))^{T}=
   \left( \begin{array}{c}\psi(0)\\
                          \psi^{\prime}(0) \end{array} \right).
\end{equation*}
Here $T$ denotes the transpose, so $[\psi],\psi(0)$ and $\psi^{\prime}(0)$ are
 considered to be column vectors of length $2n$ and $n$ respectively.
The kernel of the map $[\:]$ is obviously equal to
$\cD(\Delta^{0})$. Then we have
\begin{equation*}
\Omega(\phi,\psi)=\omega([\phi],[\psi]):=\langle[\phi],J[\psi]\rangle_{\C^{2n}},
\end{equation*}
where $\langle\,,\,\rangle_{\C^{2n}}$ now denotes the scalar product on $\C^{2n}$ and
where the $2n\times 2n$ matrix $J$ is the canonical symplectic matrix on $\C^{2n}$:
\begin{equation}
\label{1}
 J= \left ( \begin{array}{cc}0&\1\\
                            -\1&0 \end{array} \right).
\end{equation}
Here and in what follows $\1$ is the unit matrix for the given context. Note
that the Hermitian symplectic form $\omega$ differs from the Euclidean
symplectic form on $\C^{2n}$ \cite{Arnold}.

To find all maximal isotropic subspaces in $\cD$ with respect to $\Omega$ it
therefore suffices to find all maximal isotropic subspaces in $\C^{2n}$ with
respect to $\omega$ and to take their preimage under the map $[\:]$. The set of
all maximal isotropic subspaces corresponds to the Lagrangean Grassmann
manifold in the context of Euclidean symplectic forms \cite{Arnold}. Since $J$
is non degenerate, such spaces all have complex dimension equal to $n$. This
description is the local Kirchhoff rule referred to in the Introduction.
Moreover with $\cM^{\perp}$ denoting the orthogonal complement (with respect to
$\langle\cdot,\cdot\rangle_{\C^{2n}}$) of a space $\cM$ we have the
\begin{lem}
A linear subspace $\cM$ of $\C^{2n}$ is maximal isotropic iff $\cM^{\perp}=J\cM$ and iff
$\cM^{\perp}$ is maximal isotropic.
\end{lem}
The proof is standard and follows easily from the definition and the fact that
$J^{2}=-\1$ and $J^{\dagger}=-J$.
We use this result in the following form. Let the linear subspace $\cM=
\cM (A,B)$ of
$\C^{2n}$ be given as the set of all ${[\psi]}$ in $\C^{2n}$ satisfying
\begin{equation}
\label{2}
 A\psi(0)+B\psi^{\prime}(0)=0,
\end{equation}
where $A$ and $B$ are two $n\times n$ matrices. If the $n\times 2n$ matrix $(A,B)$ has
maximal rank equal to $n$ then obviously $\cM$ has dimension equal to $n$ and in this way
we may describe all subspaces of dimension equal to $n$. Also the image of $\C^{2n}$ under
the map $(A,B)$ is then all of $\C^{n}$ because of the general result that for any linear
map $T$ from $\C^{2n}$ into $\C^{n}$ one always has $\dim\:\Ker(T)+\dim\: \Ran(T)=2n$.
Writing the adjoint of any (not necessarily square) matrix $X$ as
$X^{\dagger}=\bar{X}^{T}$ we claim
\begin{lem}
Let $A$ and $B$ be be two $n\times n$ matrices such that $(A,B)$ has maximal
rank. Then $\cM(A,B)$ is maximal isotropic iff $AB^{\dagger}$ is self-adjoint.
\end{lem}
The proof is easily obtained by writing the condition \eqref{2} in the form
$\langle\Phi^{k},[\psi]\rangle_{\C^{2n}}=0$, $1\le k\le n$,  where $\Phi^{k}$ is given as
the $k$-th column vector of the $2n\times n$ matrix
$(\bar{A},\bar{B})^{T}=(A,B)^{\dagger}$. Obviously they are linearly independent.
 Then by the previous lemma $\cM(A,B)$ is maximal
isotropic iff the space spanned by the ${\Phi^{k}}$ is maximal isotropic. This condition
in turn amounts to the condition that $(A,B)J(A,B)^{\dagger}=0$, which means that
 $AB^{\dagger}$ has to be self-adjoint. The converse is also obviously true.

\begin{ex} $(n\ge 3,\;n\;odd)$  Consider
\begin{equation*}
\psi_{j}(0)+c_{j}(\psi_{j-1}^{\prime}(0)+\psi_{j+1}^{\prime}(0))
=0\;\text{for } 1\leq j\leq n
\end{equation*}
with a $\mod\: n$ convention. The resulting $A$ is the identity matrix and
 $AB^{\dagger}$ is self-adjoint iff all $c_{j}$ are equal (=c) and real.
Then $B$ is of the form
$B_{jk}=c(\delta_{j+2\;k}+\delta_{j\;k+2})$.
(If $n$ is even the condition is that $c_{j}=\bar{c}_{j+1}$ and
$c_{j}=c_{j+2}$ for all $j$)
\end{ex}

\begin{ex} $(n\ge 3)$ Consider
\begin{equation*}
\psi_{j}(0)+\psi_{j+1}(0)+c_{j}(\psi_{j}^{\prime}(0)-\psi_{j+1}^{\prime}(0))
=0\;\text{for } 1\leq j\leq n
\end{equation*}
 again with a $\mod\: n$ convention. The resulting $A$ has maximal rank and
 $AB^{\dagger}$ is self-adjoint iff now all $c_{j}$ are equal and purely
imaginary.
\end{ex}

\begin{ex} $(n=2)$ (see e.g. \cite{Seba,Carreau,Kurasov}) To relate this case to
familiar examples we realize $\cH$ as $L^{2}(\R)\;=\;L^{2}((-\infty,0])\oplus
L^{2}([0,\infty ))$ and write the boundary conditions as
\begin{equation*}
\left ( \begin{array}{c} \psi(0+)\\
                         \psi^{\prime}(0+)\end{array} \right)=
e^{i\mu}\left ( \begin{array}{cc}a&b\\
                                c&d \end{array} \right )
\left ( \begin{array}{c} \psi(0-)\\
                         \psi^{\prime}(0-)\end{array} \right).
\end{equation*}
Then $AB^{\dagger}$ is self-adjoint iff the matrix
\begin{equation*}
\left ( \begin{array}{cc}a&b\\
                                c&d \end{array} \right )
\end{equation*}
belongs to $\SL(2,\R)$ and $\mu$ is real. Up to a set of measure zero in
$\U(2)$ this gives all self-adjoint extensions. The interpretation of the
parameters entering the boundary conditions can be found in \cite{Kurasov}. The
case $a-1=d-1=b=0$, $\exp(2i\mu)
=1$ gives the familiar $\delta$-potential of strength $c$ at the origin. The
case $a-1=d-1=c=0$, $\exp(2i\mu)=1$ gives the so called $\delta^{\prime}$
-interaction of strength b (see e.g. \cite{Albeverio:book,Seba,Seba2} and references therein).
The case $a=d^{-1}$, $b=c=0$, $\exp(2i\mu)=1$ gives the $\delta^{\prime}$
-potential of strength $-2(1-a)/(1+a)$ \cite{Griffiths,Kurasov}.
In particular for the choice $a-1=d-1=b=c=0$, $\exp(2i\mu)=1$, the
corresponding free S-matrix (see below) is given by
\begin{displaymath}
S_2^\mathrm{free}(E)=\left(\begin{array}{cc}
                        0 & 1 \\
                        1 & 0 \end{array} \right).
\end{displaymath}
More generally for $n$ lines ($\simeq \R$) with free propagation and with the appropriate
labeling of the $2n$ ends the $2n\times 2n$ on-shell S-matrix takes the form,
\begin{equation}\label{s2n}
S_{2n}^\mathrm{free}(E)=\left(\begin{array}{cc}
                        0 & \1 \\
                        \1 & 0 \end{array} \right).
\end{equation}
We will make use of this observation in Section 4, where we will exploit the fact that
this $S^{free}_{2n}(E)$ serves as a unit matrix with respect to the generalized star
product.
\end{ex}

In what follows we will always assume that $A$ and $B$ define a maximal isotropic subspace
$\cM(A,B)$, such that the resulting operator, which we denote by $\Delta(A,B)$, is
self-adjoint. A core of this operator $\cD(\Delta(A,B))$ is given as the preimage of
$\cM(\Delta(A,B))$ under the map $[\:]$. Note that $\cD(\Delta^{0})$ has codimension equal
to $n$ in any of these cores. Thus the quantum mechanical  one particle Hamiltonians we
will consider are of the form $-\Delta(A,B)$ for any boundary condition $(A,B)$ defining a
maximal isotropic subspace.

The self-adjointness of $AB^{\dagger}$ , i.e. the relation $AB^{\dagger}-BA^{\dagger}=0$,
will be the main \emph{Leitmotiv} throughout this article, since combined with the maximal
rank condition it encodes the self-adjointness of the operator $\Delta(A,B)$ and is the
algebraic formulation of the local Kirchhoff rule. The proof of Lemma 2.2 combined with
the previous lemma also shows that $\cM(A,B)^{\perp}= J\cM(A,B)=\cM(-B,A)$. Note that
$(A^{\dagger},B^{\dagger})$ does not necessarily define a maximal subspace if $(A,B)$
does. As an example let $H$ be self-adjoint and $A$ invertible and set
$B=H(A^{-1})^{\dagger}$. Then $(A,B)$ defines a maximal isotropic subspace since
$AB^{\dagger}=H$, but $A^{\dagger}B=A^{\dagger}H(A^{-1})^{\dagger}$,
$B^{\dagger}A=A^{-1}HA$ and these two expressions differ if $AA^{\dagger}$ and $H$ do not
commute. If $A=0$ such that $B$ is invertible we have Neumann boundary conditions and if
$B=0$ such that $A$ is invertible we have Dirichlet boundary conditions.

We will now calculate the on-shell S-matrix $S(E)=S_{A,B}(E)$ for all energies
$E>0$. This will be an $n\times n$ matrix whose matrix elements are defined by
the following relations. We look for plane wave solutions $\psi^{k}(\cdot,E)$,
$1\le k\le n$ of the Schr\"{o}dinger equation for $-\Delta(A,B)$ in the form
\begin{equation}
\label{3}
 \psi^{k}_{j}(x,E)=\begin{cases}S_{jk}(E)e^{i\sqrt{E}x} & \text{for\;\ } j\not=k\\
                                e^{-i\sqrt{E}x}+S_{kk}(E)e^{i\sqrt{E}x}
                                          & \text{for\;\ } j=k
                   \end{cases}
\end{equation}
and which satisfy the boundary conditions. Thus $S_{kk}(E)$ has the interpretation of
being the reflection amplitude in channel $k$ while $S_{jk}(E)$ with $j\neq k$ is the
transmission amplitude from channel $k$ into channel $j$, both for an incoming plane wave
$\exp(-i\sqrt{E}x)$ in channel $k$. This definition of the S-matrix differs from the
standard one used in potential scattering theory \cite{Faddeev,Deift:Trubowitz}, where the
equal transmission amplitudes build up the diagonal. In particular for $n=2$ and general
boundary conditions at the origin as described in Example 2.3 we have
\begin{eqnarray*}
S(E)&=&\left(\begin{array}{cc}
           R(E) & T_1(E) \\
           T_2(E) & L(E)
           \end{array} \right)\\
           &=&\left(a-i\sqrt{E}b+\frac{ic}{\sqrt{E}}+d\right)^{-1}
           \left(\begin{array}{cc}
           a-i\sqrt{E}b-\frac{ic}{\sqrt{E}}-d & 2e^{i\mu} \\
           2e^{-i\mu} & -a-i\sqrt{E}b-\frac{ic}{\sqrt{E}}+d
           \end{array} \right),
\end{eqnarray*}
whereas
\begin{displaymath}
S_\mathrm{standard}(E)=S(E)\left( \begin{array}{cc}
           0 & 1 \\
           1 & 0
           \end{array}\right)=\left(\begin{array}{cc}
           T_1(E) & R(E)  \\
           L(E) & T_2(E)\end{array}\right).
\end{displaymath}
These S-matrices are unitarily equivalent iff $\Delta(A,B)$ is real (i.e. $e^{2i\mu}=1$
such that $\Delta(A,B)$ commutes with complex conjugation, see also below for a general
discussion) and is invariant with respect to reflection at the origin. In the latter case
$T_1(E)=T_2(E)$ and $R(E)=L(E)$.

We return to the Ansatz \eqref{3}. After a short calculation the boundary conditions for
the $\psi^{k}$ take the form of a matrix equation for $S(E)$
\begin{equation}
\label{4}
 (A+i\sqrt{E}B)S(E)=-(A-i\sqrt{E}B).
\end{equation}
To solve for $S(E)$ we will establish the following
\begin{lem}
For all $E>0$ both matrices $(A+i\sqrt{E}B)$ and $(A-i\sqrt{E}B)$
are invertible.
\end{lem}
Proof: Assume $\det(A+i\sqrt{E}B)=0$. But then also
\begin{equation*}
\det(A^{\dagger}-i\sqrt{E}B^{\dagger}) = \overline{\det(A+i\sqrt{E}B)}=0,
\end{equation*}
so there is $\chi\neq 0$ such that $(A^{\dagger}-i\sqrt{E}B^{\dagger})\chi=0$. In
particular
\begin{equation*}
0=\langle\chi,(A+i\sqrt{E}B)(A^{\dagger}-i\sqrt{E}B^{\dagger})\chi\rangle=
 \langle A^{\dagger}\chi,A^{\dagger}\chi\rangle+E\langle B^{\dagger}\chi,B^{\dagger}\chi\rangle,
\end{equation*}
where we have used the fact that $AB^{\dagger}$ is self-adjoint. But this implies that
$A^{\dagger}\chi=B^{\dagger}\chi =0$. Hence $\langle A\phi+B\phi^{\prime},\chi\rangle=0$
for all $\phi,\phi^{\prime}\in \C^{n}$. But as already remarked the range of the map
$(A,B):\C^{2n}\rightarrow\C^{n}$ is all of $\C^{n}$. Hence $\chi =0$ and we have arrived
at a contradiction. This proves the lemma since the invertibility of $A-i\sqrt{E}B$ is
proved in the same way. Also we have shown that $AA^{\dagger}+EBB^{\dagger}$ is a strictly
positive operator and hence an invertible operator on $\C^{n}$ for all $E>0$. If $A$ (or
$B$) is invertible there is an easier proof of the lemma. Indeed, assume there is
$\chi\neq 0$ such that $(A+i\sqrt{E}B)\chi=0$. Then we have $A^{-1}B\chi=i/\sqrt{E}\cdot
\chi$. But $A^{-1}B$ is self-adjoint since $AB^{\dagger}$ is and therefore has only real
eigenvalues giving a contradiction. To sum up we have proved the first part of
\begin{theo}
For the above quantum wire with one vertex and boundary conditions given by
the pair $(A,B)$ the on-shell S-matrix is given as
\begin{equation}
\label{5}
  \begin{array}{ccc}S_{A,B}(E)&=& -(A+i\sqrt{E}B)^{-1}(A-i\sqrt{E}B)\\
   {}&   =&-(A^{\dagger}-i\sqrt{E}B^{\dagger})(AA^{\dagger}+EBB^{\dagger})^{-1}
  (A-i\sqrt{E}B)\end{array}
\end{equation}
and is unitary and real analytic in $E>0$.
\end{theo}
To prove unitarity, we observe that
         $S^{\dagger}(E)=-(A^{\dagger}+i\sqrt{E}B^{\dagger})
          (A^{\dagger}-i\sqrt{E}B^{\dagger})^{-1}$
and $S(E)^{-1}=-(A-i\sqrt{E}B)^{-1}(A+i\sqrt{E}B)$. Now it is easy to see that these
expressions are equal using again the fact that $AB^{\dagger}$ is self-adjoint. Note that
unitarity follows from abstract reasoning. In fact the difference of the resolvents of the
operators $\Delta(A=0, B=\1)$ (Neumann boundary conditions and with S-matrix equal to
$\1$) and  $\Delta(A,B)$ by Krein's formula (see e.g. \cite{Achieser:Glasmann}) is a
finite rank operator since we are dealing with finite defect indices, so in particular
this difference is trace class. Therefore by the Birman-Kato theory the entire S-matrix
exists and is unitary. Also the S-matrix for Dirichlet boundary conditions
 $(A=\1,B=0)$ is $-\1$. The second relation in \eqref{5} can be used to discuss
the bound state problem for $-\Delta(A,B)$, which, however we will not do in
this article.

Relation \eqref{5} is a remarkable matrix analogue of the representation in potential
scattering theory of the on-shell S-matrix at given angular momentum $l$ as a quotient of
Jost functions, i.e. $S_{l}(k=\sqrt{E})=f_{l}(k)/f_{l}(-k)$ (see \cite{Jost,Newton}). As
in potential scattering theory  the scattering matrix $S_{A,B}(E)$ as a function of
$\sqrt{E}$ can be analytically continued to a meromorphic function in the whole complex plane. In fact, by the
self-adjointness of $\Delta(A,B)$ it is analytic in the physical energy sheet
$\Imt
\sqrt{E}>0$ except for poles on the positive imaginary axis corresponding to bound states and may have additional poles (i.e. resonances) in the unphysical energy
sheet $\Imt\sqrt{E}<0$.

If $C$ is any invertible $n\times n$ matrix then obviously $(CA,CB)$ defines the same
boundary conditions as $(A,B)$ such that $\cM(CA,CB)=\cM(A,B)$ and
$\Delta(CA,CB)=\Delta(A,B)$ and this is reflected by the fact that
$S_{CA,CB}(E)=S_{A,B}(E)$. Conversely, if $\cM(A,B)=\cM(A^{\prime},B^{\prime})$ then there
is an invertible $C$ such that $A=CA^{\prime},B=CB^{\prime}$. This follows easily from the
defining relation \eqref{2}. We want to use this observation to show that the on-shell
S-matrix uniquely fixes the boundary conditions. In fact assume that
$S_{A,B}(E)=S_{A^{\prime},B^{\prime}}(E)$ or equivalently
$S_{A,B}(E)S_{A^{\prime},B^{\prime}}(E)^{\dagger}=\1$ holds for some $E>0$. By a short
calculation this holds iff $A^{\prime}B^{\dagger}-B^{\prime}A^{\dagger}=0$, i.e.
$(A^\prime,B^\prime)J(A,B)^\dagger=0$. But by the proof of Lemma 2.2 this means that
$\cM(A,B)^{\perp}=\cM(A^{\prime},B^{\prime})^{\perp}$ so the two maximal isotropic
subspaces $\cM(A,B)$ and $\cM(A^{\prime},B^{\prime})$ are equal as was the claim.

To fix the freedom in parametrizing a maximal isotropic subspace by the pair
$(A,B)$, consider the Lie group $\cG(2n)$, consisting of all $2n\times 2n$
matrices $g$ which preserve the Hermitian symplectic structure, i.e.
$g^{\dagger}Jg=J$. We claim that this group is isomorphic to the classical
group $\U(n,n)$ (see e.g. \cite{He}). Indeed $iJ$ is Hermitian, $(iJ)^{2}=\1$
and $\tr\;J=0$ such that the only eigenvalues $\pm 1$ are of equal
multiplicity. Therefore there is a unitary $W$ such that
\begin{equation*}
    WiJW^{-1}= \left( \begin{array}{cc} \1&0\\
                            0&-\1 \end{array} \right).
\end{equation*}
Thus elements $g$ in the group $W\cG(2n)W^{-1}$ satisfy
\begin{equation*}
g^{\dagger}\left( \begin{array}{cc} \1&0\\
                              0&-\1 \end{array} \right)g=
\left( \begin{array}{cc} \1&0\\
                         0&-\1 \end{array} \right)
\end{equation*}
such that $W\cG(2n)W^{-1}=\U(n,n)$. The group $\U(n,n)$ and hence $\cG(2n)$ has real
dimension  $4n^{2}$ and the latter acts transitively on the set of all maximal isotropic
subspaces. In particular $\cM(A,B)$ is the image of $\cM(A=0,B=\1)$ under the map given by
the group element
\begin{equation*}
\left( \begin{array}{cc} B^{\dagger}(AA^{\dagger}+BB^{\dagger})^{-\frac{1}{2}}   &A^{\dagger}(AA^{\dagger}+BB^{\dagger})^{-\frac{1}{2}}\\
  -A^{\dagger}(AA^{\dagger}+BB^{\dagger})^{-\frac{1}{2}}
  &B^{\dagger}(AA^{\dagger}+BB^{\dagger})^{-\frac{1}{2}}\end{array} \right).
\end{equation*}
Let $\cK(2n)$ be the isotropy group of $\cM(A=0,B=\1)$, i.e. the subgproup which leaves
$\cM(A=0,B=\1)$ fixed. It is of real dimension $3n^{2}$. The set of all maximal isotropic
subspaces is in one-to-one correspondence with the right coset space $\cG(2n)/\cK(2n)$
which has dimension $n^{2}$ as it should. Also this space may be shown to be compact. In
Appendix A we will relate the present discussion of selfadjoint extensions of $\Delta^{0}$
with von Neumann's theory of selfadjoint extensions. In this context it is worthwhile to
note that the parametrization of self-adjoint extensions in terms of maximal isotropic
spaces is much more convenient for the description of these graph Hamiltonians, rather
than the standard von Neumann's parametrization. In particular the content of Appendix A
is not needed for an understanding of the main material presented in this article.

Now we establish some properties of these on-shell S-matrices. Although we shall prove
corresponding results in the general case they are more transparent and easier to prove in
this simple situation. In particular, if the boundary condition $(A,B)$ is such that $A$ is
invertible one may choose $C=A^{-1}$ and similarly $C=B^{-1}$ if $B$ is invertible. To
determine $S_{A,B}(E)$ for all $E$ in these cases it therefore suffices to diagonalize the
self-adjoint matrices $A^{-1}B$ and $B^{-1}A$ respectively. Thus if $A^{-1}B=VHV^{-1}$ with
a unitary $V$ and diagonal, self-adjoint $H$, then
$S_{A,B}(E)=-V(\1+i\sqrt{E}H)^{-1}(\1-i\sqrt{E}H)V^{-1}$.

Thus in Example 2.1 $H$ and $V$ are given as $H_{kl}=\delta_{kl}2c\cos2\pi (l-1)/n$ and
\begin{equation*}
V_{kl}=\frac{1}{\sqrt{n}}e^{\frac{2i\pi}{n}(k-1)(l-1)}
\end{equation*}
resulting in an on-shell S-matrix of the form
\begin{equation}
\label{55}
S_{jk}(E)=-\frac{1}{n}\sum_{l=1}^{n}e^{\frac{2i\pi}{n}(k-j)(l-1)}
           \frac{1-2ic\sqrt{E}\cos\frac{2\pi}{n}(l-1)}
            {1+2ic\sqrt{E}\cos\frac{2\pi}{n}(l-1)}.
\end{equation}

With the equivalence  $(A,B)\sim (CA,CB)$ for invertible $C$ in mind, it follows easily
that $S_{A,B}(E)$ is diagonal for all $E$ iff $A$ and $B$ are both diagonal (Robin
boundary conditions).

 Let $U$ be a unitary operator on $\C^{n}$. Then $U$ defines a
unitary operator $\cU$ on $\cH$ in a natural way via $(\cU
\psi)_{i}=\sum^{n}_{j=1}U_{ij}\psi_{j}$. As a special case this covers the situation where $U$ is a gauge transformation of the form $\psi_{j}\rightarrow
\exp (i\chi_{j})\psi_{j}$ with constant $\chi_{j}$.
Also $(AU,BU)$ defines a maximal isotropic
subspace and we have $\Delta(AU,BU)=\cU^{-1}\Delta(A,B)\cU$. Correspondingly we have
$S_{AU,BU}(E)=U^{-1}S_{A,B}(E)U$. Assume in particular that there is a unitary $U$ and an
invertible $C$ such that $ CA=AU$ and $CB=BU$. Then the relations $\Delta (AU,BU)=
\cU^{-1} \Delta(A,B)\cU =\Delta(A,B)$ and $S_{A,B}(E)=U^{-1}S_{A,B}(E)U$
are valid. This has a special application. There is a natural unitary representation
$\pi\rightarrow U(\pi)$ of the permutation group of $n$ elements into the unitaries of
$\C^{n}$ and analogously a representation $\pi\rightarrow\cU(\pi)$ into the unitaries of
$\cH$. For any permutation $\pi$ such that $AU(\pi)=CA$ and $BU(\pi)=CB$ holds for an
invertible $C=C(\pi)$ one has $S_{A,B}(E)=U(\pi)^{-1}S_{A,B}(E)U(\pi)$. In Examples 2.1
and 2.2 one may take $\pi$ to be the cyclic permutation and $C(\pi)=U(\pi)$. Consequently
the on-shell S-matrix for the Example 2.1 given by \eqref{55} satisfies
$S_{j+l\;k+l}(E)=S_{jk}(E)\; \mod \;n $ for all $l$ (see also e.g. \cite{Gratus} for other
examples).

Next we claim that if $A$ and $B$ are both real such that
 $A^{\dagger}=A^{T}$ and
$B^{\dagger}=B^{T}$ then $S_{A,B}(E)$ equals its transpose. In particular
the on-shell S-matrix \eqref{55} for the Example 2.1 is symmetric.
 This is in analogy
to potential scattering on the line, where the Hamiltonian is also real (i.e. commutes
with complex conjugation). There the transmission amplitude for the incoming plane wave
from the left equals the transmission amplitude for the incoming plane wave from the right
(see e.g. \cite{Faddeev,Deift:Trubowitz}). In fact one has a general result. For given
$(A,B)$ defining the self-adjoint operator $\Delta(A,B)$, $(\bar{A},\bar{B})$ also defines
a self-adjoint operator $\Delta(\bar{A},\bar{B})$, since $(\bar{A},\bar{B})$ has maximal
rank and $\bar{A}\bar{B}^{\dagger}$ is also self-adjoint. It is easy to see that
$\Delta(\bar{A},\bar{B})$ has domain $\cD(\Delta(\bar{A},\bar{B}))=\{\psi\mid
\bar{\psi}\in\cD(\Delta(A,B)\}$ and
$\Delta(\bar{A},\bar{B})\psi=\overline{\Delta(A,B)\bar{\psi}}$. In particular
$\Delta(A,B)$ is a real operator if the matrices $A$ and $B$ are real.  Thus the Laplace
operators obtained from Example 2.1 are real while those obtained from Example 2.2 are not
real. In Example 2.3 $(A,B)$ is real iff $\exp(2i\mu)=1$. More generally we will say that
the boundary conditions $(A,B)$ are real if there is an invertible map $C$ such that $CA$
and $CB$ are real. Thus Robin boundary conditions are real. For invertible $A$ (or $B$) a
necessary and sufficient condition is that the self-adjoint matrix $A^{-1}B$ (or
$B^{-1}A$) is real. We have the
\begin{cor}
The on-shell S-matrices satisfy the following relation
\begin{equation}
S_{\bar{A},\bar{B}}(E)^{T}=S_{A,B}(E).
\end{equation}
\end{cor}
For the proof observe that $S_{\bar{A},\bar{B}}(E)^{T}=
-(A^{\dagger}-i\sqrt{E}B^{\dagger})(A^{\dagger}+i\sqrt{E}B^{\dagger})^{-1}$
and the claim again easily follows from the fact that $AB^{\dagger}$ is self-adjoint.

As a next observation we want to exploit the fact that $A$ and $B$ play an almost
symmetric role. First we recall that if $(A,B)$ defines a maximal isotropic subspace then
$(-B,A)$ also defines a maximal isotropic subspace, which is just the orthogonal
complement. In particular Dirichlet boundary conditions turn into Neumann boundary
conditions and vice versa under this correspondence. Although the Laplace operators
$\Delta(A,B)$ and $\Delta(-B,A)$ are not directly related, there is a relation for their
on-shell S-matrices. In fact from
\eqref{5}
we immediately obtain
\begin{cor}
The on-shell S-matrices for the operators $-\Delta(A,B)$ and $-\Delta(-B,A)$
are related by
\begin{equation}
\label{7}
S_{-B,A}(E)=-S_{A,B}(E^{-1})
\end{equation}
for all $E>0$.
\end{cor}
Adapting the notation from string theory (see e.g. \cite{Polchinski}) we call the map
$\theta :(A,B)\mapsto (-B,A)$ a duality transformation and \eqref{7} a duality relation.
On the level of Laplace operators and on-shell S-matrices $\theta$ is obviously an
involution. Since $\cM(-B,A)=\cM(A,B)^{\perp}$ there are no selfdual boundary conditions.

We conclude this section by providing a necessary and sufficient condition for
$S_{A,B}(E)$ to be independent of $E$, i.e. to be a constant matrix. In fact this will
occur if $(A,B)$ is such that none of the linear conditions involve both $\psi(0)$ and
$\psi^{\prime}(0)$. This means that the boundary conditions are scale invariant, i.e.
invariant under the variable transformation $x\rightarrow \lambda^{-1}x$,
$x,\lambda\in\R^{+}$. The algebraic formulation is given by
\begin{cor}
Assume the boundary condition is such that to any $\lambda > 0$ there is an
invertible $C(\lambda)$ with $C(\lambda)A=A$ and $C(\lambda)B=\lambda B$. Then
both $S_{A,B}(E)$ and $S_{-B,A}(E)$ are independent of $E$. The only
eigenvalues of $S_{A,B}(E)$ are $+1$ and $-1$ with multiplicities equal to
$\Rank\;B$ and $\Rank\;A$. Also $AB^{\dagger}=0$ and both $S_{A,B}(E)$ and
$S_{-B,A}(E)$ are Hermitian such that by unitarity they are also involutive
maps. If in addition the boundary conditions are real, then $S_{A,B}(E)$ and
$S_{-B,A}(E)$ are also real. Conversely if $S_{A,B}(E)$ is constant then there
is $C(\lambda)$ with the above property.
\end{cor}
Proof: The first part is obvious. To prove the second part recall
\eqref{4}. Since $S_{A,B}(E)$ is constant this implies
$S_{A,B}(E)^{\dagger}B^{\dagger}=B^{\dagger}$ and
$S_{A,B}(E)^{\dagger}A^{\dagger}=-A^{\dagger}$. Thus the column vectors of $B^{\dagger}$
and $A^{\dagger}$ are eigenvectors of $S_{A,B}(E)^{\dagger}= S_{A,B}(E)^{-1}$ and
therefore of $S_{A,B}(E)$ with eigenvalues $+1$ and $-1$ respectively. They span subspaces
of dimensions equal to $\Rank\;B^{\dagger}= \Rank\;B$ and $\Rank\;A^{\dagger}=\Rank\;A$.
To see that these vectors combined span all of $\C^{n}$ assume there is a vector $\psi$
orthogonal to both these spaces. But this means that $A\psi=B\psi=0$ and this is possible
only if $\psi=0$ since $A-i\sqrt{E}B$ is invertible. Also eigenvectors for different
eigenvalues are orthogonal which means $AB^{\dagger}=0$. The hermiticity of $S_{A,B}(E)$
and $S_{-B,A}(E)$ therefore follows easily from
\eqref{5}. The reality of $S_{A,B}(E)$ and $S_{-B,A}(E)$ if the boundary conditions are real follows from
Corollary 2.1. As for the converse we first observe that by the above
arguments a constant $S_{A,B}(E)$ implies the previous properties concerning
eigenvalues and eigenvectors. Let therefore $U$ be a unitary map which
diagonalizes $S_{A,B}(E)$, i.e.
\begin{equation*}
  US_{A,B}(E)U^{-1}=\left ( \begin{array}{cc} -\1&0\\                                                                         0&\1 \end{array} \right)
\end{equation*}
holds with an obvious block notation.
Also trivially
\begin{equation*}
US_{A,B}(E)U^{-1}UA^{\dagger}=-UA^{\dagger},\;\;
US_{A,B}(E)U^{-1}UB^{\dagger}=UB^{\dagger}.
\end{equation*}
Therefore $A$ and $B$ are necessarily of the form
\begin{equation*}
A=\left ( \begin{array}{cc} a^{\prime}_{11}&0\\
                            a^{\prime}_{21}&0 \end{array}\right)U,\;\;
B=\left ( \begin{array}{cc} 0&b^{\prime}_{12}\\
                            0&b^{\prime}_{22} \end{array}\right)U,
\end{equation*}
again in the same block notation. Since $(A,B)$ and hence $(AU,BU)$ has maximal rank, the
matrix
\begin{equation*}
       \left( \begin{array}{cc}a^{\prime}_{11}&b^{\prime}_{12}\\
                            a^{\prime}_{21}&b^{\prime}_{22} \end{array}\right)
\end{equation*}
is invertible. Let $D$ denote its inverse, such that
\begin{equation*}
   DA=\left( \begin{array}{cc} \1&0\\
                               0&0 \end{array} \right)U,\;\;
   DB=\left( \begin{array}{cc} 0&0\\
                               0&\1 \end{array} \right)U.
\end{equation*}
Then
\begin{equation*}
  C(\lambda)=D^{-1}\left( \begin{array}{cc} \1&0\\
                                           0&\lambda\1 \end{array} \right)D
\end{equation*}
does the job, concluding the proof of the corollary.

The following example will be reconsidered in Section 4.
\begin{ex}$(n=3)$ Let the boundary conditions be given as
\begin{equation*}
\begin{array}{c}
\psi_{1}(0)=\psi_{2}(0)=\psi_{3}(0),\\[1mm]
\psi^{\prime}_{1}(0)+\psi^{\prime}_{2}(0)+\psi^{\prime}_{3}(0)=0,
\end{array}
\end{equation*}
i.e.
\begin{equation*}
 A=\left( \begin{array}{ccc} 1&-1&0\\
                             0&1&-1\\
                             0&0&0 \end{array}\right),\;
 B=\left( \begin{array}{ccc} 0&0&0\\
                             0&0&0\\
                             1&1&1 \end{array}\right),\;
 C(\lambda)=\left( \begin{array}{ccc} 1&0&0\\
                             0&1&0\\
                             0&0&\lambda \end{array}\right).
\end{equation*}
Thus $A$ has rank 2, $B$ rank 1 and $AB^{\dagger}=0$. The resulting on-shell S-matrix is
given as
\begin{equation*}
   S(E)=\left( \begin{array}{ccc} -\frac{1}{3}&\frac{2}{3}&\frac{2}{3}\\[2mm]
                                  \frac{2}{3}&-\frac{1}{3}&\frac{2}{3}\\[2mm]
                                  \frac{2}{3}&\frac{2}{3}&-\frac{1}{3}
\end{array} \right)
\end{equation*}
for all $E$. Since it is Hermitian and unitary the eigenvalues are indeed $\pm
1$ with the desired multiplicities since $\tr S(E)=-1$.
\end{ex}

\section{Arbitrary Finite Quantum Wires}

In this section we will discuss the general case employing the methods used in
the previous section. Let $\cE$ and $\cI$ be finite sets with $n$ and $m$
elements respectively and ordered in an arbitrary but fixed way. $\cE$ labels
the external lines and $\cI$ labels the internal lines, i.e. we consider a
graph with $m$ internal (of finite length) and $n$ external lines. Unless
stated otherwise $n\neq 0$, since we will mainly focus on scattering theory.
The discussion in the previous section already covered the case $m=0$ so we
will also assume $m\neq 0$. To each $e\in\cE$ we associate the infinite
interval $[0,\infty)$ and to each $i\in\cI$ the finite interval $[0,a_{i}]$
where $a_{i}>0$. The Hilbert space is now defined as
\begin{equation*}
\cH=\oplus_{e\in\cE}\cH_{e}\oplus_{i\in\cI}\cH_{i}
=\cH_{\cE}\oplus \cH_{\cI},
\end{equation*}
where $\cH_{e}=L^{2}([0,\infty))$ for $e\in\cE$ and
 $\cH_{i}=L^{2}([0,a_{i}])$ for $i\in\cI$.
$\cH_{\cE}$ is called the exterior and $\cH_{\cI}$ is called the interior component of
$\cH$. Elements in $\cH$ are written as
\begin{equation*}
\psi=\left(\{\psi_{e}\}_{e\in\cE},
\{\psi_{i}\}_{i\in\cI} \right)^T=(\psi_{\cE},\psi_{\cI})^T.
\end{equation*}
Thus the previous case is the special case when $\cI$ is empty. Let now
$\Delta^{0}$ be the Laplace  operator with domain of definition
$\cD(\Delta^{0})$ being given as the set of all $\psi$ with $\psi_{e}$,
$e\in\cE$ belongin to $W^{2,2}(0,\infty)$ and vanishing at $x=0$ together with
their first derivatives while $\psi_{i}$ belong to $W^{2,2}(0,a_{i})$ for all
$i\in\cI$ and vanish at the ends of the interval together their first
derivatives. Obviously $\Delta^{0}$ has defect indices $(n+2m,n+2m)$. To find
all self-adjoint extensions let now $\cD$ be the set of all $\psi$ with
$\psi_{e}$, $e\in\cE$ belonging to $W^{2,2}(0,\infty)$ while the $\psi_{i}\in
W^{2,2}(0,a_{i})$, $i\in\cI$. Also the skew-Hermitian quadratic form $\Omega$
on $\cD$ is now defined as
\begin{equation*}
\Omega(\phi,\psi)=\langle\Delta\phi,\psi\rangle-\langle\phi,\Delta\psi\rangle
       =-\overline{\Omega(\psi,\phi)}.
\end{equation*}
Let $[\;]:\cD\rightarrow \C^{2(n+2m)}$ be the surjective linear map which associates to
each $\psi$ the element $[\psi]$ given as
\begin{equation}\label{8}
[\psi]=\left( \begin{array}{c}(\psi_{e}(0)_{e\in\cE},
  \psi_{i}(0)_{i\in\cI},\psi_{i}(a_{i})_{i\in\cI})^{T}\\
    (\psi_{e}^{\prime}(0)_{e\in\cE},
 \psi_{i}^{\prime}(0)_{i\in\cI},
 -\psi_{i}^{\prime}(a_{i})_{i\in\cI})^{T}
           \end{array} \right)
 =\left( \begin{array}{c}\underline{\psi}\\
                  \underline{\psi}^{\prime} \end{array}\right)
\end{equation}
again viewed as a column vector with the ordering given by the ordering of $\cE$ and
$\cI$. Obviously $\cD(\Delta^{0})$ is the kernel of the map $[\;]$. By partial integration
we again obtain
\begin{equation*}
\Omega(\phi,\psi)=\omega([\phi],[\psi])=\langle[\phi],J[\psi]\rangle_{\C^{2(n+2m)}}
\end{equation*}
where now $J$ is the canonical symplectic form on $\C^{2(n+2m)}$ of the same form as in
\eqref{1} and $\langle\;,\;\rangle_{\C^{2(n+2m)}}$ is now the canonical scalar product on
$\C^{2(n+2m)}$. The formulation of the boundary condition is as in the previous section.
Let now $A$ and $B$ be $(n+2m)\times (n+2m)$ matrices and let $\cM(A,B)$ be the linear
space of all $[\psi]$ in $\C^{2(n+2m)}$ such that
\begin{equation}
\label{9}
A\underline{\psi}+B\underline{\psi}^{\prime}=0
\end{equation}
Then $\cM(A,B)$ has dimension $n+2m$ iff the $(n+2m)\times 2(n+2m)$ matrix $(A,B)$ has
maximal rank equal to $n+2m$. If in addition $AB^{\dagger}$ is self-adjoint then
$\cM(A,B)$ is maximal isotropic. The resulting self-adjoint operator will again be denoted
by $\Delta(A,B)$. Again in what follows we will always assume that the boundary conditions
$(A,B)$ have these properties.

Similar to the discussion in the case of single vertex graph (see Appendix A)
we can relate von Neumann's parametrization of self-adjoint extensions of
$\Delta^0$ to the matrices $A$ and $B$. Corresponding details can easily be
worked out and therefore we omit them.

To determine the resulting on-shell S-matrix, we now look for plane wave solutions
$\psi^{k}(\cdot,E)$, $k\in \cE$ of the stationary Schr\"odinger equation for
$-\Delta(A,B)$ at energy $E>0$ which satisfy the boundary conditions $(A,B)$ and which are
of the following form and which generalize \eqref{3}
\begin{equation}
\label{10}
\psi^{k}_{j}(x,E)=\begin{cases}S_{jk}(E)e^{i\sqrt{E}x}&\text{for}
                                          \;j\in\cE,j\neq k\\
  e^{-i\sqrt{E}x}+S_{kk}(E)e^{i\sqrt{E}x}&\text{for}\;j\in\cE,j=k\\
                                  \alpha_{jk}(E)e^{i\sqrt{E}x}+
        \beta_{jk}(E)e^{-i\sqrt{E}x}&\text{for}\;j\in\cI.\end{cases}
\end{equation}
The aim is thus to determine the $n\times n$ matrix $S(E)=S_{A,B}(E)$  and the
$m\times n$ matrices $\alpha(E)=\alpha_{A,B}(E)$ and $\beta(E)=\beta_{A,B}(E)$.
The physical interpretation of the matrix elements $S_{jk}(E)$ is as before and
the matrix elements of $\alpha (E)$ and $\beta(E)$ are `interior' amplitudes.
It is advisable to write matrices like $A$ and $B$ in a $3\times 3$ block form,
i.e.
\begin{equation*}
 A=\left(\begin{array}{ccc}A_{11}&A_{12}&A_{13}\\
                           A_{21}&A_{22}&A_{23}\\
                           A_{31}&A_{32}&A_{33}\end{array} \right),
\end{equation*}
where $A_{11}$ is an $n\times n$ matrix, $A_{12}$ and $A_{13}$ are $n\times m$
matrices etc. Thus for example the matrices
$ A_{1i}, A_{i1}, B_{1i}, B_{i1}, i=2,3$
describe the coupling of the exterior to the interior.

Correspondingly $\underline{\psi}$ and
$\underline{\psi}^{\prime}$ are written as (compare \eqref{8})
\begin{equation*}
\underline{\psi}=\left(\begin{array}{c}\psi_{\cE}(0)\\
                                       \psi_{\cI}(0)\\
                                      \psi_{\cI}(\underline{a})
                                      \end{array} \right),\
\underline{\psi}^{\prime}=\left(\begin{array}{c}\psi_{\cE}^{\prime}(0)\\
                                       \psi_{\cI}^{\prime}(0)\\
                                       -\psi_{\cI}^{\prime}(\underline{a})
                                      \end{array} \right),
\end{equation*}
where $\underline{a}=(a_{1},... a_{m})$. Also we introduce the diagonal
 $m\times m$
matrices $\exp(\pm i\sqrt{E}\underline{a})$ by
\begin{equation*}
 \exp(\pm i\sqrt{E}\underline{a})_{jk}=\delta_{jk}e^{\pm i\sqrt{E}a_{j}}\;
                       \text{for}\; j,k\in\;\cI.
\end{equation*}
The equations for the matrices $S(E),\alpha(E)$ and $\beta(E)$ now take the
following form:
\begin{eqnarray*}
 A \left( \begin{array}{c} S(E)+\1\\
    \alpha(E)+\beta(E)\\
  e^{i\sqrt{E}\underline{a}}\alpha(E)+e^{-i\sqrt{E}\underline{a}}\beta(E)
    \end{array} \right)
        +i\sqrt{E}B\left(\begin{array}{c} S(E)-\1\\
    \alpha(E)-\beta(E)\\
  -e^{i\sqrt{E}\underline{a}}\alpha(E)+e^{-i\sqrt{E}\underline{a}}\beta(E)
   \end{array} \right)=0,
\end{eqnarray*}
which is a $(n+2m)\times n$ matrix equation with matrix multiplication between
$(n+2m)\times (n+2m)$ and $(n+2m)\times n$ matrices. We rewrite this equation as an
inhomogeneous equation generalizing equation \eqref{4},
\begin{equation}
\label{11}
Z_{A,B}(E)\left( \begin{array}{c} S(E)\\
                      \alpha(E)\\
                    \beta(E)\end{array} \right) =-(A-i\sqrt{E}B)
             \left ( \begin{array}{c} \1\\
                               0\\
                               0 \end{array} \right )
\end{equation}
where
\begin{equation*}
  Z_{A,B}(E)=AX(E)+i\sqrt{E}BY(E)
\end{equation*}
with
\begin{equation*}
 X(E)= \left ( \begin{array}{ccc}\1&0&0\\
                                  0&\1&\1\\
               0&e^{i\sqrt{E}\underline{a}}&e^{-i\sqrt{E}\underline{a}}
               \end{array} \right)
\end{equation*}
and
\begin{equation*}
 Y(E)= \left ( \begin{array}{ccc}\1&0&0\\
                                  0&\1&-\1\\
               0&-e^{i\sqrt{E}\underline{a}}&e^{-i\sqrt{E}\underline{a}}
               \end{array} \right).
\end{equation*}

If  $\det Z_{A,B}(E)\neq 0$ the scattering matrix $S(E)=S_{A,B}(E)$ as well as the
$m\times n$ matrices $\alpha(E)$ and $\beta(E)$ can be uniquely determined by solving the
equation \eqref{11},
\begin{equation}\label{cont}
\left(\begin{array}{c} S(E) \\ \alpha(E) \\ \beta(E) \end{array}\right)=
-Z_{A,B}(E)^{-1}(A-i\sqrt{E}B)\left(\begin{array}{c} \1 \\ 0 \\ 0\end{array} \right).
\end{equation}
We recall that by the Birman-Kato theory $S_{A,B}(E)$ is defined and unitary for almost
all $E>0$ because $\Delta(A,B)$ is a finite rank perturbation of $\Delta(A=0,B=\1)$. We
denote by $\Sigma_{A,B}=\left\{E>0\ | \det Z_{A,B}(E)=0 \right\}$ the set of exceptional
points for which $Z_{A,B}(E)$ is not invertible. Now we prove
\begin{theo}
For any boundary condition $(A,B)$ the set $\Sigma_{A,B}$ equals the set $\sigma_{A,B}$ of
all positive eigenvalues of $-\Delta(A,B)$. This set is discrete and has no finite
accumulation points in $\R_+$.
\end{theo}

Proof: First we prove that for every $E\in\Sigma_{A,B}$ all elements of $\Ker Z_{A,B}(E)$
have the form $(0,\widehat{\alpha}^{T},\widehat{\beta}^{T})^T$ with $\widehat{\alpha}$ and
$\widehat{\beta}$ being (column) vectors in $\C^m$. Let us suppose the converse, i.e. that
there is a column vector
$(\widehat{s}^{T},\widehat{\alpha}^{T},\widehat{\beta}^{T})^T\in\C^{n+2m}$ such that
\begin{equation}\label{null}
Z_{A,B}(E)\left(\begin{array}{c}\widehat{s} \\ \widehat{\alpha} \\ \widehat{\beta}
\end{array}\right)=0,
\end{equation}
or equivalently
\begin{displaymath}
(A+i\sqrt{E}B)\left(\begin{array}{c} \widehat{s} \\ \widehat{\alpha} \\
e^{-i\sqrt{E}\underline{a}}\widehat{\beta}
\end{array}\right)+(A-i\sqrt{E}B)\left(\begin{array}{c} 0 \\ \widehat{\beta} \\
e^{i\sqrt{E}\underline{a}}\widehat{\alpha}
\end{array}\right)=0.
\end{displaymath}
As in the proofs of Lemma 2.3 and Theorem 2.1  $A+i\sqrt{E}B$ and $A-i\sqrt{E}B$ are
invertible and $(A+i\sqrt{E}B)^{-1}(A-i\sqrt{E}B)$ is unitary such that
\begin{displaymath}
\left(\begin{array}{c} \widehat{s} \\ \widehat{\alpha} \\
e^{-i\sqrt{E}\underline{a}}\widehat{\beta}
\end{array}\right)=-(A+i\sqrt{E}B)^{-1}(A-i\sqrt{E}B)\left(\begin{array}{c} 0 \\ \widehat{\beta} \\
e^{i\sqrt{E}\underline{a}}\widehat{\alpha}
\end{array}\right).
\end{displaymath}
Since unitary transformations preserve the Euclidean norm we get
\begin{displaymath}
\|\widehat{s}\|^2_{\C^n}+\|\widehat{\alpha}\|^2_{\C^m}+\|\widehat{\beta}\|^2_{\C^m}=
\|\widehat{\alpha}\|^2_{\C^m}+\|\widehat{\beta}\|^2_{\C^m},
\end{displaymath}
such that $\widehat{s}=0$.

Now for arbitrary $(0,\widehat{\alpha}^{T},\widehat{\beta}^{T})^{T}\in\Ker Z_{A,B}(E)$ we consider
\begin{displaymath}
\psi_j(x)=\left\{\begin{array}{lll}
              0 & \text{for} & j\in\cE,\\
              \widehat{\alpha}_je^{i\sqrt{E}x}+\widehat{\beta}_je^{-i\sqrt{E}x} & \text{for} &
              j\in\cI. \end{array}\right.
\end{displaymath}
Obviously $\psi(x)$ is an eigenfunction of $-\Delta(A,B)$. Thus we have proved
that $\Sigma_{A,B}\subseteq\sigma_{A,B}$. We note that this inclusion is
nontrivial, since a priori we cannot exclude real energy resonances.

Conversely, let $E\in\sigma_{A,B}$. Observing that positive energy eigenfunctions must
have support on the internal lines of the graph and repeating the arguments, which led to
equation \eqref{11}, we see that there exists a nonzero vector $(0,\widehat{\alpha}^{T},
\widehat{\beta}^{T})^T\in\C^{n+2m}$ such that \eqref{null} is satisfied, and hence
$\sigma_{A,B}\subseteq\Sigma_{A,B}$. We notice that $\det Z_{A,B}(E)$ is an entire
function of $\sqrt{E}$ in the complex plane $\C$. Also $\det Z_{A,B}(E)$ does not vanish
identically since by the preceding arguments $\det Z_{A,B}(E)=0$ for $E\in\C\setminus\R$
implies that $E$ is an eigenvalue of $-\Delta(A,B)$, which in turn contradicts the
self-adjointness. Thus all real zeroes of $\det Z_{A,B}(E)$ are isolated. This concludes
the proof of the theorem.

\begin{re}
The case $n=0$ with no associated S-matrix is of interest in its own right. Then
\eqref{11} takes the form of a homogeneous equation
\begin{equation}
\label{111}
Z_{A,B}(E)\left( \begin{array}{c}\widehat{\alpha}(E)\\
                           \widehat{\beta}(E) \end{array} \right)=0,
\end{equation}
with $\widehat{\alpha}(E)$ and $\widehat{\beta}(E)$ being column vectors in $\C^m$. It has
solutions iff $E\in
\R$ is such that $\det\;Z_{A,B}(E)=0$. By what has been said so far it is
clear that the solutions of
\eqref{111} give the eigenvalues and the eigenfunctions of a quantum wire without open
ends. As a possible example one might consider the graph associated to the semiclassical
description of the fullerene and choose appropriate boundary conditions at the vertices
(see e.g. \cite{Fullerene}).
\end{re}

Before we proceed further with the study of the equation \eqref{11} we consider several
examples. Starting with the case, where everything works well without any singularities,
we consider the following
\begin{ex}$(n=m=1)$ We choose the following realization of the Hilbert space $\cH$:
\begin{equation*}
\cH\;=\;L^{2}([0,\infty ))\;=\;L^{2}([1,\infty))\oplus L^{2}([0,1]).
\end{equation*}
We take Robin boundary conditions at the origin and a $\delta$ potential of
strength $c$ at $x=1$, i.e.
\begin{equation*}
\begin{array}{cccc} \sin\varphi\;\psi(0)+\cos\varphi\;\psi^{\prime}(0)&=&0&{}\\
 \psi(1+)-\psi(1-)&=&0&{}\\
 \psi^{\prime}(1+)-\psi^{\prime}(1-)-c\psi(1)&=&0&. \end{array}
\end{equation*}
$S(E),\;\alpha(E)$ and $\beta(E)$ are now functions and a straightforward calculation
gives the following result. Let $S_{R}(E)$ denote the on-shell S-matrix for the Robin
boundary condition alone, i.e.
\begin{equation*}
S_{R}(E)=e^{2i\delta_{R}(E)}=-\frac{\sin\varphi-i\sqrt{E}\cos\varphi}
                                   {\sin\varphi+i\sqrt{E}\cos\varphi}.
\end{equation*}
Then
\begin{equation*}
\begin{array}{ccc} S(E)&=&
 \frac{(2i\sqrt{E}+c)e^{i(\sqrt{E}+2\delta_{R}(E))}+ce^{-i\sqrt{E}}}
 {(2i\sqrt{E}-c)e^{-i\sqrt{E}}-ce^{i(\sqrt{E}+2\delta_{R}(E))}}e^{-2i\sqrt{E}}\\
      \\
 \alpha(E)&=&\frac{2i\sqrt{E}e^{-i(\sqrt{E}-2\delta_{R}(E))}}
             {(2i\sqrt{E}-c)e^{-i\sqrt{E}}+ce^{i(\sqrt{E}+2\delta_{R}(E))}}\\
       \\
 \beta(E)&=&\frac{2i\sqrt{E}e^{-i\sqrt{E}}}
           {(2i\sqrt{E}-c)e^{-i\sqrt{E}}+ce^{i(\sqrt{E}+2\delta_{R}(E))}}.
     \end{array}
\end{equation*}
In particular these quantities are finite for all $E>0$.
\end{ex}

\begin{figure}[ht]
\centerline{
\unitlength1mm
\begin{picture}(120,60)
\put(40,30){\circle*{2}}
\put(80,30){\circle*{2}}
\put(10,30){\line(1,0){30}}
\put(80,30){\line(1,0){30}}
\put(60,30){\oval(40,40)}
\put(30,30){\vector(-1,0){10}}
\put(90,30){\vector(1,0){10}}
\put(25,31){1}
\put(95,31){2}
\put(60,10){\vector(1,0){5}}
\put(60,50){\vector(1,0){5}}
\put(60,6){4}
\put(60,51){3}
\end{picture}}
\caption{\label{fig:1} The graph from Example 3.2. The arrows show the positive direction
for every segment.}
\end{figure}
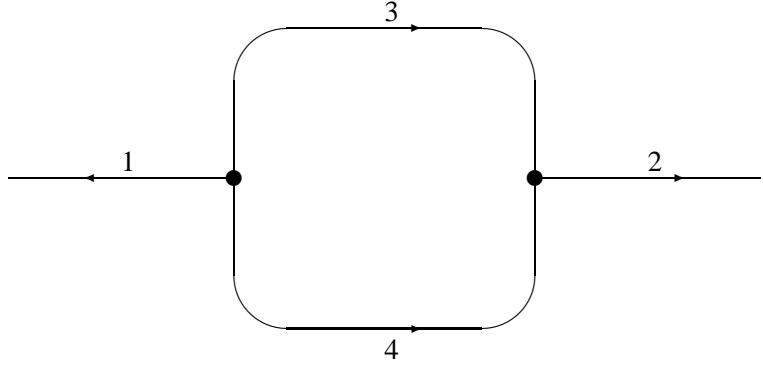

\begin{ex} Consider the graph depicted in Fig. \ref{fig:1} with
$\cE=\{1,2\}$, $\cI=\{3,4 \}$, and
with $a_{3}=a_{4}=a$, i.e. the internal lines have equal length.
 The arrows on the graph segments show the positive
directions. We pose the following real boundary conditions, obviously defining a
self-adjoint operator,
\begin{eqnarray*}
\psi_1(0)=\psi_3(0)=\psi_4(0),\\
\psi_2(0)=\psi_3(a)=\psi_4(a),\\
\psi'_1(0)+\psi'_3(0)+\psi'_4(0)=0,\\
\psi'_2(0)-\psi'_3(a)-\psi'_4(a)=0.
\end{eqnarray*}
Straightforward calculations yields
\begin{equation*}
\det Z(E)=(10-e^{2i\sqrt{E}a}-9\,e^{-2i\sqrt{E}a})E
\end{equation*}
such that $\Sigma_{A,B}=\{\frac{n^2\pi^2}{a^{2}},\ n\in\N\}$. The corresponding wave
functions have the form $\psi_1=\psi_2\equiv 0$, $\psi_3(x)=-\psi_4(x)=\sin(\frac{n\pi
x}{a})$. Note that these eigenfunctions for the embedded eigenvalues have compact support.
This is in contrast to standard Schr\"{o}dinger operators of the form $-\Delta\,+V$, where
bound state eigenfunctions can not have compact support due to the ellipticity of
$-\Delta$. The on-shell S-matrix can also be easily calculated, giving
\begin{equation}\label{pearl}
S(E)=-\frac{1}{e^{2i\sqrt{E}a}-9}
\left( \begin{array}{cc} 3(e^{2i\sqrt{E}a}-1)&8e^{i\sqrt{E}a}\\
                          8e^{i\sqrt{E}a}&3(e^{2i\sqrt{E}a}-1)
         \end{array} \right).
\end{equation}
Equation \eqref{11} is solvable for all  $E\in\Sigma_{A,B}$ and defines $S(E)$
uniquely. $S(E)$ is well behaved for all $E>0$ and there is no influence of the
bound states, except that the (equal) reflection amplitudes vanish for
$E\in\Sigma_{A,B}$. This is in contrast to Schr\"odinger operators of the form
$-\Delta+\lambda P$ with appropriate real $\lambda$ and with $P$ being the
orthogonal projector onto any given one-dimensional linear subspace of the
Hilbert space \cite{Bolsterli}, or to Schr\"odinger operators with Wigner --
von Neumann type potentials \cite{Klaus}. The scattering matrix \eqref{pearl}
has poles only in the unphysical sheet (resonances),
\begin{displaymath}
\sqrt{E_n}=\frac{\pi n}{a}-i\frac{\log 9}{2a},\ n\in\Z.
\end{displaymath}
The relation
\begin{equation*}
\lim_{a\downarrow 0}S(E)=\left( \begin{array}{cc}0&1\\
                       1&0 \end{array}\right)=
                       S_{2}^\mathrm{free}(E)
\end{equation*}
holds as to be expected from the boundary conditions.
\end{ex}

Let us observe that
\begin{equation*}
 \det\;X(E)=\det\;Y(E)=\prod_{j\in\cI}(-2i\sin\sqrt{E}a_{j}).
\end{equation*}
Therefore if $E\in \Sigma_{\underline{a}}
=\cup_{j\in\cI}\Sigma_{\underline{a}}(j)
=\cup_{j\in\cI}\{ E> 0\mid \sin\sqrt{E}a_{j}=0\}$ then
$Z_{A,B}(E)$ will not be invertible for $(A,B)$ defining Dirichlet or Neumann boundary
conditions, since then $\det\;Z_{A,B}(E)=\det\;X(E)$. In particular in these two cases the
exterior and the interior decouple such that then we have
\begin{equation*}
\Delta(A,B)=\Delta_{\cE}(A,B)\oplus\Delta_{\cI}(A,B)
\end{equation*}
Here $\Delta_{\cE}(A,B)$ for both $(A=\1,B=0)$ and $(A=0,B=\1)$ have an absolutely
continuous spectrum and $\Delta_{\cI}(A,B)$
 has a purely discrete spectrum
which on the set $E>0$ equals $\Sigma_{\underline{a}}$. This means that we have
eigenvalues embedded in the continuum. Now the equation
\eqref{11} for Dirichlet or Neumann boundary conditions has a unique solution
$S(E)=-\1$ and $S(E)=\1$, respectively, and $\alpha(E)=\beta(E)=0$ whenever $E$ is not in
$\Sigma_{\underline{a}}$. If $E$ is in $\Sigma_{\underline{a}}$, then $S(E)$ is still of
this form but $\alpha(E)$ and $\beta(E)$ are nonunique and of the form
$\alpha(E)=\beta(E)$ and $\alpha(E)=-\beta(E)$ for Dirichlet and Neumann boundary
conditions, respectively, with $\alpha_{jk}$ being arbitrary whenever
$E\in\Sigma_{\underline{a}}(j)$ and zero otherwise.

A similar result  is valid for arbitrary boundary conditions:

\begin{theo} For any boundary condition $(A,B)$
and all $E\in\Sigma_{A,B}$ the equation \eqref{11} is solvable and determines $S_{A,B}(E)$
uniquely.
\end{theo}

Proof: Let $\psi$ be an arbitrary element of $\Ker\ Z(E)^\dagger$, i.e. $\psi$ solves the
adjoint homogeneous equation corresponding to \eqref{11},
\begin{equation}\label{adjung}
(X(E)^\dagger A^\dagger-i\sqrt{E}Y(E)^\dagger B^\dagger)\psi=0.
\end{equation}
Suppose first that $B=0$, such that $A$ has full rank (Dirichlet boundary conditions).
Moreover $\Sigma_{A,B}=\Sigma_{\underline{a}}$. Any solution of \eqref{adjung} may be
written as $\psi={A^\dagger}^{-1}\chi$ with $\chi$ being a solution of
$X(E)^\dagger\chi=0$. $\chi$ is necessarily of the form
$\chi=(0,\widehat{\alpha}^{T},\widehat{\beta}^{T})^T$ for some
$\widehat{\alpha},\widehat{\beta}\in\C^m$. Let $\phi_i\in\C^{n+2m}$, $i=1,\ldots,n$ be
such that $(\phi_i)_k=\delta_{ik}$. Then
\begin{displaymath}
(\psi, A\phi_i)=(A^\dagger\psi,\phi_i)=(\chi,\phi_i)=0
\end{displaymath}
for all $i=1,\ldots,n$. We now apply the Fredholm alternative as follows. First we note
that $\Ker\,Z(E)^{\dagger}=(\Ran \,Z(E))^{\perp}$, the orthogonal complement of the range
of $Z(E)$. Therefore the last relation states that all column vectors of
\begin{equation*}
(A-i\sqrt{E}B)\left( \begin{array}{c}\1\\
                                  0\\
                                  0\end{array}\right)
                              =A\left( \begin{array}{c}\1\\
                                  0\\
                                  0\end{array}\right)
\end{equation*}
are in the range of $Z(E)$. Thus \eqref{11} has a solution and since all elements of
$\Ker\,Z(E)$ are of the form $(0,\widehat{\alpha}^{T},\widehat{\beta}^{T})^{T}$, the
on-shell S-matrix $S_{A,B}(E)$ is determined uniquely.

Now we suppose that $B\neq 0$. First we consider the case $E\in\Sigma_{A,B}$ and
$E\notin\Sigma_{\underline{a}}$. From \eqref{adjung} it follows that
\begin{displaymath}
A^\dagger\psi=i\sqrt{E}{X(E)^\dagger}^{-1}Y(E)^\dagger B^\dagger\psi,
\end{displaymath}
and thus
\begin{equation}\label{adjung1}
(\psi, BA^\dagger \psi)=i\sqrt{E}(B^\dagger\psi, {X(E)^\dagger}^{-1}Y(E)^\dagger
B^\dagger\psi).
\end{equation}
Since $BA^\dagger$ is self-adjoint the l.h.s. of \eqref{adjung1} is real. To analyze the
r.h.s. we note that
\begin{equation}\label{form}
{X(E)^\dagger}^{-1}Y(E)^\dagger=\left(\begin{array}{ccc}
           \1 & 0 & 0\\
           0 & 0 & 0 \\
           0 & 0 & 0 \end{array}\right)-i\left(\begin{array}{ccc}
           0 & 0 & 0 \\
           0 & \mathrm{cotan}(\sqrt{E}\underline{a}) & -\frac{1}{\sin(\sqrt{E}\underline{a})}\\
           0 & -\frac{1}{\sin(\sqrt{E}\underline{a})} & \mathrm{cotan}(\sqrt{E}\underline{a})
           \end{array}\right),
\end{equation}
where the first term is self-adjoint and the second skew-self-adjoint. Therefore
\begin{displaymath}
\Ret(B^\dagger\psi, {X(E)^\dagger}^{-1}Y(E)^\dagger B^\dagger\psi)=
\left(B^\dagger\psi,\left(\begin{array}{ccc}
           \1 & 0 & 0\\
           0 & 0 & 0 \\
           0 & 0 & 0 \end{array}\right) B^\dagger\psi\right).
\end{displaymath}
Since the l.h.s. of \eqref{adjung1} is real, it follows that
\begin{equation}\label{again}
\left(B^\dagger\psi,\left(\begin{array}{ccc}
           \1 & 0 & 0\\
           0 & 0 & 0 \\
           0 & 0 & 0 \end{array}\right) B^\dagger\psi\right)=0,
\end{equation}
and thus
\begin{equation}\label{step1}
(\phi_i, B^\dagger\psi)=0
\end{equation}
for all $i=1,\ldots,n$. Multiplying the equation \eqref{adjung} by $\phi_i$ from the left
we obtain
\begin{displaymath}
(\phi_i,A^\dagger\psi)-i\sqrt{E}(\phi_i, B^\dagger\psi)=0.
\end{displaymath}
Thus from \eqref{step1} it follows that $(\phi_i,A^\dagger\psi)=0$. Therefore
\begin{displaymath}
(\psi,(A-i\sqrt{E}B)\phi_i)=0
\end{displaymath}
for all $i=1,\ldots,n$. Therefore we may again invoke the Fredholm alternative and obtain
a unique on-shell S-matrix. Finally we turn to the case
$E\in\Sigma_{A,B}\cap\Sigma_{\underline{a}}$. An important ingredient of the proof in this
case is the Moore--Penrose generalized inverse (or pseudoinverse) (see e.g.
\cite{Penrose}). Recall that for any (not necessary square) matrix
$M$ its generalized inverse $M^\star$ is uniquely defined by the Penrose equations
\begin{eqnarray*}
M M^\star M =M, && M^\star M M^\star = M^\star,\\ (M^\star M)^\dagger=M^\star M, && (M
M^\star)^\dagger= M M^\star.
\end{eqnarray*}
Also one has
\begin{eqnarray*}
(M^\star)^\dagger &=& (M^\dagger)^\star,\\
\Ran M^\star &=& \Ran M^\dagger,\\
\Ker M^\star &=& \Ker M^\dagger,
\end{eqnarray*}
and $M M^\star=P_{\Ran M}$, $M^\star M=P_{\Ran M^\dagger}$, where $P_{\cH}$ denotes the
orthogonal projector onto the subspace $\cH$. Moreover, $0^\star=0$. If $M$ is a square
matrix of full rank $M^\star=M^{-1}$. However, the product formula for inverse matrices
$(M_1 M_2)^\star=M_2^\star M_1^\star$ does not hold in general. The pseudoinverse of any
diagonal  matrix $\Lambda$ with $\det\ \Lambda=0$ is given by
\begin{displaymath}
\Lambda^\star=\left(\begin{array}{cccccc}
\lambda_1 &        & & & & \\
          & \ddots & & & & \\
          &        & \lambda_r & & & \\
          &        & & 0 & & \\
          &        & & & \ddots  & \\
          &        & & & &  0 \end{array}\right)^\star=
          \left(\begin{array}{cccccc}
\lambda_1^{-1} &        & & & & \\
          & \ddots & & & & \\
          &        & \lambda_r^{-1} & & & \\
          &        & & 0 & & \\
          &        & & & \ddots  & \\
          &        & & & &  0 \end{array}\right).
\end{displaymath}
Using the fact that $(Q^\dagger M Q)^\star=Q^\dagger M^\star Q$ for any unitary $Q$ one
can easily calculate the generalized inverse by means of the formulas
\begin{displaymath}
M^\star=(M^\dagger M)^\star M^\dagger= M^\dagger (MM^\dagger)^\star.
\end{displaymath}

With these preparatory remarks we return to the proof. We will say that $\chi\in \Ker X(E)^\dagger$ is a basis element of
$\Ker X(E)^\dagger$ if $\chi=(0,\widehat{\alpha}^{T},\widehat{\beta}^{T})^{T}$ with
$\widehat{\alpha}_i=\widehat{\beta}_i=0$ for all $i=1,\ldots,m$ except for some $k=1,\ldots,m$
and $\widehat{\alpha}_k\neq 0$, $\widehat{\beta}_k\neq 0$. For any basis element $\chi$
either
\begin{equation}\label{case1}
X(E)\chi=0\qquad\mbox{and}\qquad Y(E)\chi=c_Y\chi
\end{equation}
or
\begin{equation}\label{case2}
Y(E)\chi=0\qquad\mbox{and}\qquad X(E)\chi=c_X\chi
\end{equation}
with some $c_X, c_Y\neq 0$. The proof is elementary and is left to the reader. Taking the scalar product of \eqref{adjung} with $\chi$  we obtain
\begin{displaymath}
(X(E)\chi, A^\dagger\psi)=i\sqrt{E}(Y(E)\chi, B^\dagger\psi).
\end{displaymath}
In the case \eqref{case1} we have $(\chi,B^\dagger\psi)$=0, and in the case \eqref{case2}
$(\chi,A^\dagger\psi)=0$.

Multiplying equation \eqref{adjung} by ${X(E)^\star}^\dagger$ we obtain
\begin{displaymath}
P_{\Ran X(E)}A^\dagger\psi=i\sqrt{E}{X(E)^\star}^\dagger Y(E)^\dagger B^\dagger\psi,
\end{displaymath}
and thus
\begin{equation}\label{adjung3}
(\psi, BP_{\Ran X(E)}A^\dagger \psi)=i\sqrt{E}(B^\dagger\psi, {X(E)^\star}^\dagger
Y(E)^\dagger B^\dagger\psi).
\end{equation}
The matrix ${X(E)^\star}^\dagger Y(E)^\dagger$ has the form of the r.h.s. of
\eqref{form}, where the singular entries must be replaced by zeroes. Thus
\begin{displaymath}
\Ret(B^\dagger\psi, {X(E)^\star}^\dagger Y(E)^\dagger B^\dagger\psi)=
\left(B^\dagger\psi,\left(\begin{array}{ccc}
           \1 & 0 & 0\\
           0 & 0 & 0 \\
           0 & 0 & 0 \end{array}\right) B^\dagger\psi\right).
\end{displaymath}
Using the relation $P_{\Ran\,X(E)}=\1\,-P_{\Ran\,X(E)^{\perp}}=
\1\,-P_{\Ker\,X(E)^{\dagger}}$ we may rewrite the l.h.s. of \eqref{adjung3} as
\begin{displaymath}
(\psi, BP_{\Ran X(E)}A^\dagger\psi)=(\psi, BA^\dagger\psi)- (B^\dagger\psi, P_{\Ker
X(E)^\dagger}A^\dagger\psi).
\end{displaymath}
With $\chi_i\,(1\le i\le \dim\, \Ker\,X(E)^{\dagger})$ being the basis elements of $\Ker
X(E)^\dagger$ we have
\begin{displaymath}
(B^\dagger\psi, P_{\Ker X(E)^\dagger}A^\dagger\psi)=\sum_{i=1}^{\dim\Ker X(E)^\dagger}
\overline{(\chi_i, B^\dagger\psi)}(\chi_i, A^\dagger\psi).
\end{displaymath}
By the discussion above all terms in this sum are zero. Thus we again obtain
\eqref{again}. The rest of the proof is as in the preceding cases.
Theorem 3.2 in particular says that the presence of bound states does not spoil the
existence and uniqueness of the on-shell S-matrix. This is to be expected since bound
states do not participate in the scattering. The same is true for resonances due to their
finite lifetime.

We are now prepared to formulate the main result of this article
\begin{theo}
For any boundary condition $(A,B)$ defining the self-adjoint operator
$-\Delta(A,B)$ the resulting on-shell S-matrix $S_{A,B}(E)$ is unitary for all
$E\in\R_{+}$.
\end{theo}
Proof: The proof is quite elementary. By arguments used in the
proof of Theorem 3.1 we obtain that any solution of
\eqref{11} satisfies
\begin{displaymath}
\left(\begin{array}{c} S(E) \\ \alpha(E) \\
e^{-i\sqrt{E}\underline{a}}\beta(E)
\end{array}\right)=-(A+i\sqrt{E}B)^{-1}(A-i\sqrt{E}B)\left(\begin{array}{c} \1 \\ \beta(E) \\
e^{i\sqrt{E}\underline{a}}\alpha(E)
\end{array}\right).
\end{displaymath}
Now $(A+i\sqrt{E}B)^{-1}(A-i\sqrt{E}B)$ is unitary. Multiplying each side with its adjoint from the left we therefore obtain
\begin{equation*}
 S(E)^{\dagger}S(E)+\alpha(E)^{\dagger}\alpha(E)+\beta(E)^{\dagger}\beta(E)
=\1+\beta(E)^{\dagger}\beta(E)+\alpha(E)^{\dagger}\alpha(E)
\end{equation*}
which gives $S(E)^{\dagger} S(E)=\1$ and which is unitarity. A second
`analytic' proof is given in Appendix B (see also Section 4 for a third proof
based on the generalized star product).

We now discuss some properties of $S_{A,B}(E)$. Obviously
$Z_{A,B}(E)$ can be analytically continued to the complex $\sqrt{E}$-plane. By the
self-adjointness of $\Delta(A,B)$ the determinant $\det Z_{A,B}(E)$ cannot have zeroes for
$E$ with $\Imt\sqrt{E}>0$ except for those on the positive imaginary semiaxis
corresponding to the bound states. It may have additional zeroes (resonances) in the
unphysical energy sheet $\Imt\sqrt{E}<0$. By means of \eqref{cont} the scattering matrix
$S_{A,B}(E)$ can be analytically continued to the whole complex plane as a meromorphic
function of $\sqrt{E}$. In fact it is a rational function in $\sqrt{E}$,
$\exp(i\sqrt{E}a_{j})$, and $\exp(-i\sqrt{E}a_{j})$. Therefore it is real analytic for all
$E\in\R_+\setminus\Sigma_{A,B}$. Since $S_{A,B}(E)$ for $E\in\Sigma_{A,B}$ is well defined
and unitary, it is also continuous for all $E\in\R$.

As for the low and high energy behaviour we have the following obvious property. If $A$ is
invertible  then $\lim_{E\downarrow 0} S_{A,B}(E)=-\1$ and if $B$ is invertible then
$\lim_{E\uparrow \infty} S_{A,B}(E)=\1$. For arbitrary $(A,B)$ the corresponding relations
do not hold in general as may be seen from looking at the Dirichlet $(A=\1, B=0)$ and
Neumann $(A=0, B=\1)$ boundary conditions.

Let $C$ be an invertible map on $\C^{n+2m}$ such that
$\Delta(CA,CB)=\Delta(A,B)$. Correspondingly we have $S_{CA,CB}(E)=S_{A,B}(E)$
as it should be, since $Z_{CA,CB}(E)=CZ_{A,B}(E)$.
Furthermore let $U$ be a unitary map on $\C^{n}$. Then $U$ induces a unitary
map $\hat{U}$ on $\C^{n+2m}$ by
\begin{equation*}
 \hat{U}=\left( \begin{array}{ccc} U&0&0\\
                            0&\1&0\\
                            0&0&\1\end{array} \right)
\end{equation*}
and a unitary $\cU$ on $\cH$ via
\begin{equation*}
  (\cU\psi)_{j}=\begin{cases}\sum_{k\in\cE}U_{jk}\psi_{k}\text{ for }
                               j\in\cE\\
                  \psi_{j} \text{ otherwise } \end{cases},
\end{equation*}
such that $\Delta(A\hat{U},B\hat{U})=\cU^{-1}\Delta(A,B)\cU$. We recall that all spaces
$\cH_{e}\,(e\in\cE)$ are the $L^{2}$ space $L^{2}([0,\infty ))$, so the definition of
$\cU$ makes sense. Also $Z_{A\hat{U},B\hat{U}}(E)=Z_{A,B}(E)\hat{U}$ since $\hat{U}$
commutes with $X(E)$ and $Y(E)$, such that $\Sigma_{A,B}=\Sigma_{A\hat{U},B\hat{U}}$. Next
we observe that
\begin{equation*}
\hat{U}\left( \begin{array}{c} S_{A\hat{U},B\hat{U}}(E)\\
                                \alpha_{A\hat{U},B\hat{U}}(E)\\
                  \beta_{A\hat{U},B\hat{U}}(E)\end{array} \right)
   =\left( \begin{array}{c}US_{A\hat{U},B\hat{U}}(E)\\
                                \alpha_{A\hat{U},B\hat{U}}(E)\\
                  \beta_{A\hat{U},B\hat{U}}(E)\end{array} \right)
\end{equation*}
and
\begin{equation*}
(A\hat{U}-i\sqrt{E}B\hat{U})\left( \begin{array}{c} \1\\
                                                   0\\
     0\end{array} \right)
   =(A-i\sqrt{E}B)\left( \begin{array}{c} \1\\
                                       0\\
                                     0 \end{array} \right)U.
\end{equation*}
From this we immediately obtain first for $E$ outside $\Sigma_{A,B}$ and then by
continuity for all $E>0$

\begin{cor}
 The following covariance properties hold for
 all $E>0$
\begin{equation}
\begin{array}{ccc} S_{A\hat{U},B\hat{U}}(E)&=&U^{-1}S_{A,B}(E)U\\
                   \alpha_{A\hat{U},B\hat{U}}(E)&=&\alpha_{A,B}(E)U\\
           \beta_{A\hat{U},B\hat{U}}(E)&=&\beta_{A,B}(E)U.\end{array}
\end{equation}
 In particular if $U$ is such that there exists an invertible $C=C(U)$ with
$CA=A\hat{U}$ and $CB=B\hat{U}$ then $S_{A,B}(E)=U^{-1}S_{A,B}(E)U$ for all $E>0.$
\end{cor}
We have the following special application. There is a canonical representation
$\pi\rightarrow U(\pi)$ of the permutation group of $n$ elements into the unitaries of
$\C^{n}$. If there is a $\pi$ and an invertible $C=C(\pi)$ such that $CA=A\hat{U}(\pi)$
and $CB=B\hat{U}(\pi)$ then $S_{A,B}(E)=U^{-1}(\pi)S_{A,B}(E)U(\pi)$ for all $E>0$. The
on-shell S-matrix in Example 3.2 is obviously invariant under the permutation
$1\leftrightarrow2$ of the two external legs.

\begin{re} This discussion may be extended to the case that some of the
interval lengths $a_{i}$ are equal resulting in more general covariance and
possibly invariance properties described by some $U$, $\cU$ and $\hat{U}$
such that $\hat{U}_{11}=U$ and
$\hat{U}_{i1}=\hat{U}_{1i}=0$ for $i=2,3$. We leave out the details, which may be worked out easily.
\end{re}

Corollary 2.1 has the following generalization
\begin{cor}
For all boundary conditions $(A,B)$ and all $E>0$
 the following relation holds
\begin{equation}
      S_{\bar{A},\bar{B}}(E)^{T}=S_{A,B}(E).
\end{equation}
In particular for real boundary conditions the transmission coefficient from channel $j$
to channel $k,\;k\neq j\;(j,k\in \cE)$ equals the transmission coefficient from channel
$k$ to channel $j$ for all $E>0$.
\end{cor}
Proof: We start with two remarks. First the relation
\begin{equation*}
 (A-i\sqrt{E}B) \left( \begin{array}{c}\1\\
                                       0\\
                                      0 \end{array} \right)
=(AX^{\prime}-i\sqrt{E}BY^{\prime})
  \left( \begin{array}{c}\1\\
                                       0\\
                                      0 \end{array} \right)
\end{equation*}
holds for any $X^{\prime}$  of the form
\begin{equation*}
   X^{\prime}=\left( \begin{array}{ccc}\1&0&0\\
                                       0&X^{\prime}_{22}&X^{\prime}_{23}\\
                               0&X^{\prime}_{32}&X^{\prime}_{33}
                     \end{array} \right)
\end{equation*}
and similarly for $Y^{\prime}$. Secondly the matrix
\begin{equation*}
         \left(  \begin{array}{c} S_{A,B}(E)\\
                                  \alpha_{A,B}(E)\\
                                 \beta_{A,B}(E)
                \end{array} \right)
\end{equation*}
constitutes the first $n$ columns of the matrix
$-Z_{A,B}(E)^{-1}(AX^{\prime}-i\sqrt{E}BY^{\prime}B)$ with $X^{\prime}$ and
$Y^{\prime}$ arbitrary as above. Assume for a moment that $E$ is not
in $\Sigma_{\underline{a}}$. To prove the corollary for such $E$ it therefore
suffices to show that
\begin{eqnarray*}
(Z_{\bar{A},\bar{B}}(E)^{-1}(\bar{A}Y(E)^{-1^{T}}
-i\sqrt{E}\bar{B}X(E)^{-1^{T}}))^{T}\\
=Z_{A,B}(E)^{-1}(AY(E)^{-1^{T}}-i\sqrt{E}BX(E)^{-1^{T}}),
\end{eqnarray*}
which is equivalent to
\begin{eqnarray*}
(AX(E)+i\sqrt{E}BY(E))(Y(E)^{-1}A^{\dagger}-i\sqrt{E}X(E)^{-1}B^{\dagger})\\
=(AY(E)^{-1^{T}}-i\sqrt{E}BX(E)^{-1^{T}})
(X(E)^{T}A^{\dagger}+i\sqrt{E}Y(E)^{T}B^{\dagger}).
\end{eqnarray*}
But this relation follows from the self-adjointness of $AB^{\dagger}$
 and the observation that $X(E)Y(E)^{-1}$ is symmetric, such that the
relations $X(E)Y(E)^{-1}=Y(E)^{-1^{T}}X(E)^{T}$ and $Y(E)X(E)^{-1}=X(E)^{-1^{T}}Y(E)^{T}$
hold. Finally by continuity we may drop the condition that $E$ is not in
$\Sigma_{\underline{a}}$, thus completing the proof. The on-shell S-matrix
of Example 3.2 is obviously symmetric.

To generalize the duality map of the previous section, we have to take the
$\underline{a}$ dependence into account, so we write $S_{A,B,\underline{a}}$
etc. The reason is that the length scales $a_{j}$ induce corresponding
energy scales.
Also the Hilbert spaces depend on $\underline{a}$,
$\cH=\cH_{\underline{a}}$, so we will compare on-shell S-matrices related to
theories in different Hilbert spaces.

For given $\underline{a}$ let $\underline{a}(E)$ be given by $a_{i}(E)=Ea_{i}, i\in\cI$
such that $X_{\underline{a}(E)}(E^{-1})=X_{\underline{a}}(E)$ and
$Y_{\underline{a}(E)}(E^{-1})=Y_{\underline{a}}(E)$ for all $E>0$. Also set
\begin{equation*}
T= \left( \begin{array}{ccc} \1&0&0\\
                             0&\1&0\\
                             0&0&-\1 \end{array} \right),
\end{equation*}
such that $T^{\dagger}=T,\;T^{2}=\1$ and
$ TX_{\underline{a}}(E)T=Y_{\underline{a}}(E)$. We now define $\theta(A,B)
=(-BT,AT)$. It is easy to see that $\theta(A,B)$ defines a maximal subspace.
Also $E^{-1}$ is not in $\Sigma_{\theta(A,B),\underline{a}(E)}$ if $E$
is not in
$\Sigma_{A,B,\underline{a}}$ and vice versa.

This leads to following generalization of Corollary 2.2
\begin{cor}
For all boundary conditions (A,B) and all $E>0$ the following identities hold
\begin{equation}
\begin{array}{cccc}
 S_{\theta(A,B),\underline{a}(E)}(E^{-1})&=&-&S_{A,B,\underline{a}}(E)\\
 \alpha_{\theta (A,B),\underline{a}(E)}(E^{-1})&=
     &-&\alpha_{A,B,\underline{a}}(E)\\
 \beta_{\theta (A,B),\underline{a}(E)}(E^{-1})&=
&{}&\beta_{A,B,\underline{a}}(E).
\end{array}
\end{equation}
\end{cor}
The proof is easy by observing that
\begin{equation*}
Z_{\theta(A,B),\underline{a}(E)}(E^{-1})
=\frac{i}{\sqrt{E}}Z_{A,B,\underline{a}}(E)T
\end{equation*}
and
\begin{equation*}
-BT-\frac{i}{\sqrt{E}}AT=-\frac{i}{\sqrt{E}}(A-i\sqrt{E}B)T.
\end{equation*}

We conclude this section by giving a geometrical description of an arbitrary boundary
condition as a local boundary condition at the vertices of a suitable graph.

For given sets $\cE$ and $\cI$ and $\underline{a}$, we label the halfline $[0,\infty )$
associated to $e\in\cE$ by $I_{e}=[0_{e},\infty_{e})$ and the closed interval $[0,a_{i}]$
associated to $i\in\cI$ by $I_{i}=[0_{i},a_{i}]$ (considering $a_{i}$ as a generic
variable there should be no confusion between the number $a_{i}$ and the label $a_{i}$).
By $I=\cup_{e\in\cE}I_{e}\cup_{i\in\cI}I_{i}$ we denote the disjoint union. Let
$\cV=\cup_{e\in\cE}\{0_{e}\}\cup_{i\in\cI}\{0_{i},a_{i}\}
\subset I$ be the set of `endpoints' in $I$. Clearly the number of points in
$\cV$ equals $\mid\cE\mid +2\mid\cI\mid= n+2m$. Consider a decomposition
\begin{equation}
\label{15}
\cV=\cup_{\xi\in\Xi}\cV_{\xi}
\end{equation}
of $\cV$ into nonempty disjoint subsets $\cV_{\xi}$ with $\Xi$ being just an
index set. We say that the points in $\cV\subset I$ are equivalent ($\sim$)
when they lie in the same $\cV_{\xi}$. By identifying equivalent points in
$\cV\subset I$ we obtain a graph $\Gamma$, $\Gamma=I/\sim$. In mathematical
language $\Gamma$ is a one-dimensional simplicial complex, which in particular
is a topological space and noncompact if $\cE$ is nonempty. Obviously the
vertices in $\Gamma$ are in one-to-one correspondence with the elements
$\xi\in\Xi$. Note that $\Gamma$ need not be connected. Also there may be
`tadpoles', i.e. we allow that $0_{i}$ and $a_{i}$ for some
 $i\in\cI$ belong to a same set $\cV_{\xi}$.
There is no restriction on the number of lines entering a vertex. In particular this
number may equal 1 (so called dead end side branches \cite{Alex}), see Example 3.1. The
graph need not be planar.

Let $\{A,B\}$ be the equivalence class of the boundary condition $(A,B)$ with respect to
the equivalence relation given as $ (A^{\prime},B^{\prime})\sim (A,B)$ iff there exists an
invertible $C$ such that $A^{\prime}=CA$, $B^{\prime}=CB$. By our previous discussion
$\Delta(A,B)$ only depends on $\{A,B\}$. We say that $\{A,B\}$ has a description as a
local boundary condition on the graph $\Gamma$ if the following holds. First observe that
we may label the columns of $A$ and $B$ by the elements $v$ in $\cV$ (see \eqref{8}). With
this convention there is supposed to exist $(A^{\prime},B^{\prime})\in\{A,B\}$ with the
following properties. To each $k$ labeling the rows of $A^{\prime}$ and $B^{\prime}$ there
is $\xi=\xi (k)$, such that $A_{kv}^{\prime}=B_{kv}^{\prime}=0$ for all $v$ not in
$\cV_{\xi}$. In other words the boundary condition labeled by $k$ only involves the value
of $\psi$ and its derivative at those points in $\cV$ which belong to $\cV_{\xi}$ and this
set is in one-to-one correspondence with a vertex in $\Gamma$. Of course if $\Gamma$ is
the unique graph consisting of one vertex only then this $\Gamma$ does the job for any
boundary condition $\{A,B\}$. However, one may convince oneself that for any given
boundary condition $\{A,B\}$ there is a unique maximal graph $\Gamma=\Gamma(\{A,B\})$
describing $\{A,B\}$ as a local boundary condition, where maximal means that the number of
vertices is maximal.

Let us briefly indicate the proof. Arrange the
 $2(n+2m)$ columns
of the $(n+2m)\times 2(n+2m)$ matrix $(A,B)$ in such a way that the first $n+2m$ columns
are linearly independent. Call this matrix $X$. Then there is an invertible matrix $C$
such that the $(n+2m)\times (n+2m)$ matrix made of the first $n+2m$ columns of $CX$ is the
unit matrix. Now rearrange $CX$ by undoing the previous arrangement giving $(CA,CB)$.The
decomposition
\eqref{15} and hence $\Gamma(\{A,B\})$ may be read off the `connectivity' of $(CA,CB)$.
The self-adjointness of $AB^{\dagger}$ may now be verified locally at each vertex (the
local Kirchhoff rule), see Examples 3.1 and 3.2. Of course different boundary conditions
may still give the same $\Gamma$ in this way. Also in this sense the graphs associated to the
discussion in Section 2 may actually consist of several disconnected parts, each with one
vertex but without tadpoles.

In particular this discussion shows that the results of this section
cover all local boundary conditions on all graphs with arbitrary lengths in
the interior and which describe Hamiltonians with free propagation away from
the vertices.

\section{The Generalized Star Product and Factorization of the S-matrix}

In this section we will define a new composition rule for unitary matrices not necessarily
of equal rank. This composition rule will generalize the star product for unitary $2\times
2$ matrices so we will call it a generalized star product. It will be associative and the
resulting matrix will again be unitary. We will apply this new composition rule to obtain
the on-shell S-matrix for a graph from the on-shell S-matrices at the same energy of two
subgraphs obtained by cutting the graph along an arbitrary numbers of lines. By iteration
this will in particular allow us to obtain the on-shell S-matrix for an arbitrary graph
from the on-shell S-matrices associated to its vertices (see Section 2), thus leading to a
third proof of unitarity.

Let $V$ be any unitary $p\times p$ matrix $(p>0)$. The composition rule will depend on $V$
and will be denoted by $*_{V}$, such that for any unitary $n^{\prime}\times n^{\prime}$
matrix $U^{\prime}$ with $n^{\prime}\ge p$ and any unitary $n^{\prime\prime}\times
n^{\prime\prime}$ matrix $U^{\prime\prime}$ with $n^{\prime\prime}\ge p,\;2p<
n^{\prime}+n^{\prime\prime}$ and subject to a  certain condition (see below) there will be
a resulting unitary $n\times n$ matrix $U=U^{\prime}*_{V}U^{\prime\prime}$ with
$n=n^{\prime}+ n^{\prime\prime}-2p$. This generalized star product may be viewed as an
amalgamation of $U^{\prime}$ and $U^{\prime\prime}$ and with $V$ acting as an amalgam. To
construct $*_{V}$ we write $U^{\prime}$ and $U^{\prime\prime}$ in a $2\times 2$-block form
\begin{equation}
\label{54}
U^{\prime}=\left ( \begin{array}{cc} U^{\prime}_{11}&U^{\prime}_{12}\\
                        U^{\prime}_{21}&U^{\prime}_{22}\end{array} \right),\
U^{\prime\prime}
=\left ( \begin{array}{cc} U^{\prime\prime}_{11}&U^{\prime\prime}_{12}\\
       U^{\prime\prime}_{21}&U^{\prime\prime}_{22}\end{array} \right),
\end{equation}
where $U^{\prime}_{22}$ and $U^{\prime\prime}_{11}$ are $p\times p$ matrices,
$U^{\prime}_{11}$ is an $(n^{\prime}-p)\times (n^{\prime}-p)$ matrix,
$U^{\prime\prime}_{22}$ is an
$(n^{\prime\prime}-p)\times (n^{\prime\prime}-p)$ matrix etc. The unitarity
condition for $U^{\prime}$ then reads
\begin{equation*}
\begin{array}{ccc} U^{\prime^{\dagger}}_{11}U^{\prime}_{11}+
  U^{\prime^{\dagger}}_{21}U^{\prime}_{21}&=&\1\\[1mm]
U^{\prime^{\dagger}}_{12}U^{\prime}_{12}+
  U^{\prime^{\dagger}}_{22}U^{\prime}_{22}&=&\1\\[1mm]
U^{\prime^{\dagger}}_{11}U^{\prime}_{12}+
  U^{\prime^{\dagger}}_{21}U^{\prime}_{22}&=&0\\[1mm]
U^{\prime^{\dagger}}_{12}U^{\prime}_{11}+
  U^{\prime^{\dagger}}_{22}U^{\prime}_{21}&=&0 \end{array}
\end{equation*}
and similarly for $U^{\prime\prime}$.

\vspace{2mm}

\underline{\textit{Condition A}}: The $p\times p$ matrix
$VU^{\prime}_{22}V^{-1}U^{\prime\prime}_{11}$ does not have 1 as an eigenvalue.

\vspace{2mm}

Note that by unitarity of $U^{\prime},\,U^{\prime\prime}$ and $V$ one has
$\parallel VU_{22}^{\prime}V^{-1}U_{11}^{\prime\prime}\parallel\le1$.
Strict inequality holds whenever $\parallel U^{\prime}_{22}\parallel
 <1$ or $\parallel U_{11}^{\prime\prime}\parallel <1$ and then $Condition\;A$
 is satisfied.
In general if \textit{Condition A} is satisfied it is easy to see that the following $p\times p$
matrices exist.
\begin{equation*}
\begin{array}{ccc} K_{1}&=&(\1-VU^{\prime}_{22}V^{-1}U^{\prime\prime}_{11})^{-1}V
=V(1-U^{\prime}_{22}V^{-1}U^{\prime\prime}_{11}V)^{-1},\\
&&\\
K_{2}&=&(\1-V^{-1}U^{\prime\prime}_{11}VU^{\prime}_{22})^{-1}V^{-1}
=V^{-1}(1-U^{\prime\prime}_{11}VU^{\prime}_{22}V^{-1})^{-1}.\end{array}
\end{equation*}
An easy calculation establishes the following relations
\begin{equation}
\label{345}
\begin{array}{ccc}
K_{1}&=&V+VU^{\prime}_{22}V^{-1}U^{\prime\prime}_{11}K_{1}=
V+VU^{\prime}_{22}K_{2}U^{\prime\prime}_{11}V\\
&&\\
&=&V+K_{1}U^{\prime}_{22}V^{-1}U^{\prime\prime}_{11}V,\\
&&\\
K_{2}&=&V^{-1}+V^{-1}U^{\prime\prime}_{11}VU^{\prime}_{22}K_{2}=
V^{-1}+V^{-1}U^{\prime\prime}_{11}K_{1}U^{\prime}_{22}V^{-1}\\
&&\\
      &=&V^{-1}+K_{2}U^{\prime\prime}_{11}VU^{\prime}_{22}V^{-1}.
\end{array}
\end{equation}
Note that formally one has
\begin{equation}
\label{555}
\begin{array}{ccc}
K_{1}&=&\displaystyle\sum^{\infty}_{m=0}
   (VU^{\prime}_{22}V^{-1}U^{\prime\prime}_{11})^{m}V,\\[2mm]
K_{2}&=&\displaystyle\sum^{\infty}_{m=0}
   (V^{-1}U^{\prime\prime}_{11}VU^{\prime}_{22})^{m}V^{-1}.
\end{array}
\end{equation}
With these preparations the matrix $U=U^{\prime}*_{V}U^{\prime\prime}$ is now
defined as follows. Write $U$ in a $2\times 2$ block form as
\begin{equation*}
        U=\left( \begin{array}{cc}U_{11}&U_{12}\\
                                  U_{21}&U_{22} \end{array} \right),
\end{equation*}
where $U_{11}$ is an $(n^{\prime}-p)\times (n^{\prime}-p)$ matrix,
$U_{22}$ is an $(n^{\prime\prime}-p)\times (n^{\prime\prime}-p)$ matrix etc.
These matrices are now defined as
\begin{equation}
\label{56}
\begin{array}{ccc}
 U_{11}&=&U^{\prime}_{11}
        +U^{\prime}_{12}K_{2}U^{\prime\prime}_{11}VU^{\prime}_{21},\\
 &&\\
 U_{22}&=&U^{\prime\prime}_{22}
      +U^{\prime\prime}_{21}K_{1}U^{\prime}_{22}V^{-1}U^{\prime\prime}_{12},\\
  &&\\
 U_{12}&=&U^{\prime}_{12}K_{2}U^{\prime\prime}_{12},\\
  &&\\
 U_{21}&=&U^{\prime\prime}_{21}K_{1}U^{\prime}_{21}. \end{array}
\end{equation}
In particular if $n^\prime=2p$ then
\begin{displaymath}
\left(\begin{array}{cc}
0 & \1 \\ \1 & 0 \end{array}\right)*_V U^{\prime\prime}=
\left(\begin{array}{cc}
V^{-1} & 0 \\ 0 & \1 \end{array}\right)U^{\prime\prime}
\left(\begin{array}{cc}
V & 0 \\ 0 & \1 \end{array}\right).
\end{displaymath}
Similarly if $n^{\prime\prime}=2p$ then
\begin{displaymath}
U^\prime*_V\left(\begin{array}{cc} 0 & \1 \\ \1 & 0 \end{array}\right)=
\left(\begin{array}{cc}
\1 & 0 \\ 0 & V \end{array}\right)U^\prime\left(\begin{array}{cc}
\1 & 0 \\ 0 & V^{-1} \end{array}\right).
\end{displaymath}
In this sense the matrices $\left(\begin{array}{cc} 0 & \1 \\ \1 & 0 \end{array}\right)$
serve as units when $V=\1$.

A straightforward but somewhat lengthy calculation presented in Appendix C gives
\begin{theo}
If \textit{Condition A} is satisfied then the matrix $U=U^{\prime}*_{V}U^{\prime\prime}$
is unitary.
\end{theo}
Analogously one may prove associativity. More precisely let $U^{\prime\prime\prime}$ be a
unitary $n^{\prime\prime\prime}\times n^{\prime\prime\prime}$ and $V^\prime$ a unitary
$p^\prime\times p^\prime$ matrix with $p^\prime\leq n^{\prime\prime}$, $p^\prime\leq
n^{\prime\prime\prime}$. If $p+p^\prime\leq n^\prime$, then
\begin{equation*}
U^{\prime}*_{V}(U^{\prime\prime}*_{V^{\prime}}U^{\prime\prime\prime})=
(U^{\prime}*_{V}U^{\prime\prime})*_{V^{\prime}}U^{\prime\prime\prime}
\end{equation*}
holds whenever \textit{Condition A} is satisfied for the compositions involved.

We apply this to the on-shell S-matrices of quantum wires as follows. For the special case
$V=\1$ we introduce the notation $*_{p}=*_{V=\1}$. Let $\Gamma^{\prime}$ and
$\Gamma^{\prime\prime}$ be two graphs with $n^{\prime}$ and $n^{\prime\prime}$ external
lines labeled by $\cE^{\prime}$ and $\cE^{\prime\prime}$, i.e.
$|\cE^{\prime}|=n^{\prime}$, $|\cE^{\prime\prime}|=n^{\prime\prime}$ and an arbitrary
number of internal lines. Furthermore at all vertices we have local boundary conditions
giving Laplace operators $\Delta(\Gamma^{\prime})$ on $\Gamma^{\prime}$ and
$\Delta(\Gamma^{\prime\prime})$ on $\Gamma^{\prime\prime}$ and on-shell S-matrices
$S^{\prime}(E)$ and $S^{\prime\prime}(E)$. Let now $\cE^{\prime}_{0}$ and
$\cE^{\prime\prime}_{0}$ be subsets of $\cE^{\prime}$ and $\cE^{\prime\prime}$
respectively having an equal number $(=p>0)$ of elements. Also let $\varphi_{0}:\;
\cE^{\prime}_{0}\rightarrow
\;\cE^{\prime\prime}_{0}$ be a one-to-one map. Finally to each $k\in
\cE^{\prime}_{0}$ we associate a number $a_{k}>0$. With these data
we can now form a graph $\Gamma$ by connecting the external line $k\in\cE^{\prime}_{0}$
with the line $\varphi_{0}(k)\in\cE^{\prime\prime}_{0}$ to form a line of length $a_{k}$.
In other words the intervals $[0_{k},\infty_{k})$ belonging to $\Gamma^{\prime}$ and the
intervals $[0_{\varphi_{0}(k)},\infty_{\varphi_{0}(k)})$ belonging to
$\Gamma^{\prime\prime}$ are replaced by the the finite interval $[0_{k},a_{k}]$ with
$0_{k}$ being associated to the same vertex in $\Gamma^{\prime}$ as previously and $a_{k}$
being associated to the same vertex in $\Gamma^{\prime\prime}$ as $0_{\phi_{0}(k)}$ before
in the sense of the discussion at the end of Section 3. Recall that the graphs need not be
planar. Thus $\Gamma$ has $n=n^{\prime}+n^{\prime\prime}-2p$ external lines indexed by
elements in $(\cE^{\prime}\setminus\cE^{\prime}_{0})\cup(\cE^{\prime\prime}\setminus
\cE^{\prime\prime}_{0})$ and $p$ internal lines indexed by elements in $\cE^{\prime}_{0}$
in addition to those of $\Gamma^{\prime}$ and $\Gamma^{\prime\prime}$. There are no new
vertices in addition to those of $\Gamma^{\prime}$ and $\Gamma^{\prime\prime}$ so the
boundary conditions on $\Gamma^{\prime}$ and $\Gamma^{\prime\prime}$ define boundary
conditions on $\Gamma$ resulting in a Laplace operator $\Delta(\Gamma)$. The following
formula relates the corresponding on-shell S-matrices $S^{\prime}(E)$,
$S^{\prime\prime}(E)$ and $S(E)$. First let the indices of $\cE^{\prime}_{0}$ in
$\cE^{\prime}$ come after the indices in $\cE^{\prime}\setminus\cE^{\prime}_{0}$ (in an
arbitrary but fixed order) (see \eqref{54}). Via the map $\varphi_{0}$ we may identify
$\cE^{\prime\prime}_{0}$ with $\cE^{\prime}_{0}$ so let these indices now come first in
$\cE^{\prime\prime}$, but again in the same order. Finally let the diagonal matrix
$V(\underline{a})$ be given as
\begin{displaymath}
V(\underline{a})=\left( \begin{array}{cc}
        \exp i\sqrt{E}\underline{a}&0\\
        0&\1 \end{array}\right),
\end{displaymath}
where $\exp(i\sqrt{E}\underline{a})$ again is the diagonal $p\times p$ matrix given by the
$p$ lengths $a_{k}$, $k\varepsilon \cE^{\prime}_{0}$. Then we claim that the relation
\begin{equation}\label{compos}
S(E)=S^\prime(E)*_p V(\underline{a})S^{\prime\prime}(E)V(\underline{a})
\end{equation}
holds. The operators $K_{1}$ and $K_{2}$ involved now depend on $E$ and are singular for
$E$ in a denumerable set, namely when $Condition\;A$ is violated. This follows by
arguments similar to those made after Theorem 3.3. Thus these values have to be left out
in \eqref{compos}. In the end one may then extend
\eqref{compos} to these singular values of $E$ by arguments analogous to those after Remark 3.1.
If $\Gamma$ is simply the disjoint union of $\Gamma^{\prime}$ and $\Gamma^{\prime\prime}$,
i.e. if no connections are made (corresponding to $p=0$ and
$n=n^{\prime}+n^{\prime\prime}$), then $S(E)$ is just the tensor product of
$S^{\prime}(E)$ and $S^{\prime\prime}(E)$. In this sense the generalized star product is a
generalization of the tensor product. Also by a previous discussion (see (\ref{s2n}))
$V^{-1}S(E)V=S_{2n}^\mathrm{free}(E)*_V S(E)$ for any on-shell S-matrix with $n$ open ends
and any unitary $n\times n$ matrix $V$. Similarly $S(E)*_V
S_{2n}^\mathrm{free}(E)=VS(E)V^{-1}$. Using \eqref{compos} the on-shell S-matrix
associated to any graph and its boundary conditions is obtained from the on-shell
S-matrices associated to its subgraphs each having one vertex only. In fact, pick one
vertex and choose all the internal lines connecting to all other vertices. This leads to
two graphs and the rule \eqref{compos} may be applied. Iterating this procedure $L$ times,
where $L$ is the number of vertices, gives the desired result.
\begin{figure}[ht]
\centerline{
\unitlength1mm
\begin{picture}(120,60)
\put(40,30){\circle{14}}
\put(80,30){\circle{14}}
\put(40,30){$\Gamma^\prime$}
\put(80,30){$\Gamma^{\prime\prime}$}
\put(47.1,30){\line(1,0){26}}
\put(47,32){\line(1,0){26}}
\put(47,28){\line(1,0){26}}
\put(33,30){\line(-1,0){10}}
\put(21,30){\line(-1,0){2}}
\put(17,30){\line(-1,0){2}}
\put(87,30){\line(1,0){10}}
\put(99,30){\line(1,0){2}}
\put(103,30){\line(1,0){2}}
\put(33.5,33){\line(-1,1){7}}
\put(25.5,41){\line(-1,1){4}}
\put(20.5,46){\line(-1,1){4}}
\put(86.5,33){\line(1,1){7}}
\put(94.5,41){\line(1,1){4}}
\put(99.5,46){\line(1,1){4}}
\put(33.5,27){\line(-1,-1){7}}
\put(25.5,19){\line(-1,-1){4}}
\put(20.5,14){\line(-1,-1){4}}
\put(86.5,27){\line(1,-1){7}}
\put(94.5,19){\line(1,-1){4}}
\put(99.5,14){\line(1,-1){4}}
\end{picture}}
\caption{\label{fig:2} Decomposition of the graph.}
\end{figure}
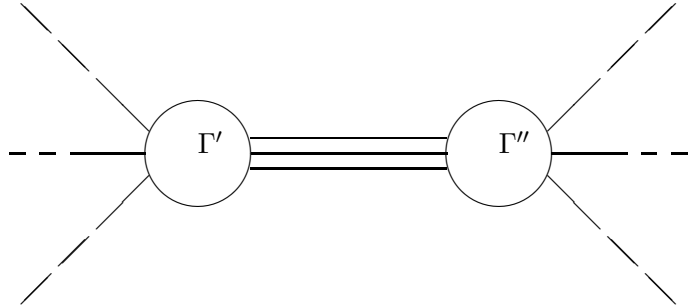

We will not prove the claim \eqref{compos} here but only give formal and intuitive
arguments which also apply in the physical context of the usual star product or Aktosun
formula for potential scattering on the line and which are based on a rearrangement of the
Born series for the on-shell S-matrix. For partial results in potential scattering in
higher dimensions when the separation of two (or more) potentials tends to infinity see
\cite{Cluster1,Cluster2}. The complete proof of \eqref{compos} will be given elsewhere
\cite{elsewhere}.

For the sake of definiteness we consider the amplitudes $S(E)_{kl}$,
$k,l\in\cE^{\prime}\setminus\cE^{\prime}_{0}$, which form $U_{11}$ in this context, since
we have $U^{\prime}=S^{\prime}(E)$ etc. The other amplitudes may be discussed analogously.
The first nontrivial contribution is $S^{\prime}(E)_{kl}$ corresponding to the first term
in the first relation of
\eqref{56}. In next order the incoming plane wave in channel $l$ may cross within
$\Gamma^{\prime}$ into channel $k^{\prime}\in\cE^{\prime}_{0}$ picking up a factor
$S^{\prime}(E)_{k^{\prime}l}$. Then it propagates from $\Gamma^{\prime}$ to
$\Gamma^{\prime\prime}$ picking up a phase factor $\exp(i\sqrt{E}a_{k^{\prime}})$. To end
up in channel $k$ it then has to be transmitted within $\Gamma^{\prime\prime}$ into
another channel $k^{\prime\prime}$ which is in $\cE^{\prime\prime}_{0}$ and which we have
identified with $\cE^{\prime}_{0}$. Therefore it picks up a factor
$S^{\prime\prime}(E)_{k^{\prime\prime}\;k^{\prime}}$ and then a factor
$\exp(i\sqrt{E}a_{k^{\prime\prime}})$ when propagating back to $\Gamma^{\prime}$ and
finally comes the factor $S^{\prime}(E)_{k\;k^{\prime\prime}}$ from propagation within
$\Gamma^{\prime}$ before ending up in channel $k$. By the superposition principle
summation has to be carried over all such $k^{\prime}$ and $k^{\prime\prime}$. This
contribution therefore corresponds to the term with $m=0$ in the expression for $K_{2}$ in
\eqref{555} when inserted into the second term in the first relation of
\eqref{56}. The higher order contributions arise if
the plane wave is reflected $m+1>1$ times back and forth between $\Gamma^{\prime}$ and
$\Gamma^{\prime\prime}$. Again by the superposition principle one finally has to sum over
all $m$, giving the desired relation for $S(E)_{kl}$.

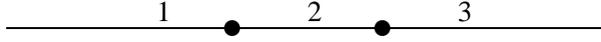
\begin{figure}[ht]
\centerline{
\unitlength1mm
\begin{picture}(120,40)
\put(20,20){\line(1,0){80}}
\put(50,20){\circle*{2}}
\put(70,20){\circle*{2}}
\put(40,21){1}
\put(60,21){2}
\put(80,21){3}
\end{picture}}
\caption{\label{fig:line} The graph from Example 4.1.}
\end{figure}

\begin{ex}
Consider an arbitrary self-adjoint Laplacian $\Delta(A,B)$ with local boundary
conditions on the graph depicted in Fig. \ref{fig:line}, where the distance
between the vertices is $a$. The composition rule (\ref{compos}) with
\begin{displaymath}
V(a)=\left(\begin{array}{cc} e^{i\sqrt{E}a} & 0 \\ 0 & 1 \end{array}\right)
\end{displaymath}
easily gives
\begin{eqnarray*}
S_{11} &=& S^\prime_{11}+S^\prime_{12}S_{11}^{\prime\prime}S^\prime_{21}
(1-S^\prime_{22}S^{\prime\prime}_{11}e^{2ia\sqrt{E}})^{-1},\\
S_{22} &=&S^{\prime\prime}_{22}+ S_{22}^\prime
S_{21}^{\prime\prime}S_{12}^{\prime\prime}
(1-S^\prime_{22}S^{\prime\prime}_{11}e^{2ia\sqrt{E}})^{-1},\\
S_{12} &=& S_{12}^\prime
S_{12}^{\prime\prime}(1-S^\prime_{22}S^{\prime\prime}_{11}e^{2ia\sqrt{E}})^{-1},\\
S_{21}
&=&
S_{21}^{\prime\prime}S_{21}^\prime(1-S^\prime_{22}S^{\prime\prime}_{11}e^{2ia\sqrt{E}})^{-1},
\end{eqnarray*}
where the S-matrices are written in the form analogous to \eqref{54}
\begin{displaymath}
S^\prime = \left(\begin{array}{cc}S^\prime_{11} & S^\prime_{12} \\
                    S^{\prime}_{21} & S_{22}^\prime\end{array}\right),\
S^{\prime\prime} = \left(\begin{array}{cc}S^{\prime\prime}_{11} &
S^{\prime\prime}_{12} \\
                    S^{\prime\prime}_{21} & S_{22}^{\prime\prime}\end{array}\right),
\end{displaymath}
leaving out the $E-$dependence. These relations are equivalent to the Aktosun
factorization formula applied to the Laplacian on a line with boundary
conditions posed at $x=0$ and $x=a$.
\end{ex}

\begin{ex}
Consider the Laplacian on the graph from Example 3.2 (Fig. \ref{fig:1}). The
relation (\ref{compos}) holds with
\begin{displaymath}
V(a)=\left(\begin{array}{ccc} e^{i\sqrt{E}a} & 0 & 0 \\
                          0 & e^{i\sqrt{E}a} & 0 \\
                          0 & 0 & 1 \end{array}\right)
\end{displaymath}
and $S^\prime=S^{\prime\prime}$ calculated in the Example 2.4. It is easy to check that
\begin{displaymath}
K_1=K_2=(1-e^{2i\sqrt{E}a}/9)^{-1}(1-e^{2i\sqrt{E}a})^{-1}
\left(\begin{array}{cc} 1-\frac{5}{9}e^{2i\sqrt{E}a} & -\frac{4}{9}e^{2i\sqrt{E}a}\\
-\frac{4}{9}e^{2i\sqrt{E}a} & 1-\frac{5}{9}e^{2i\sqrt{E}a}\end{array} \right).
\end{displaymath}
Therefore from (\ref{56}) the formula (\ref{pearl}) again follows. Note that $K_{1}=K_{2}$
is singular at $\Sigma_{A,B}$, but these singularities disappear in the on-shell S-matrix.
\end{ex}

As already remarked multiple application of (\ref{compos}) to an arbitrary
graph allows one by complete induction on the number of vertices to calculate
its S-matrix from the S-matrices corresponding to single-vertex graphs. If
these single vertex graphs contain no tadpoles, i.e. internal lines starting
and ending at the vertex, then (\ref{5}) and (\ref{compos}) give a complete
explicit construction of the S-matrix in terms of the on-shell S-matrices
discussed in Section 2. In case when a resulting single-vertex graph contains
tadpoles we proceed as follows. Let the graph $\Gamma$ have one vertex, $n$
external lines and $m$ tadpoles of lengths $a_i$. To calculate the S-matrix of
$\Gamma$ we insert an extra vertex on each of the internal lines (for
definiteness, say, at $x=a_{i}/2$). At these new vertices we impose trivial
boundary conditions given by the choice $a-1=d-1=b=c=0,\; \exp(2i\mu) =1$ of
Example 3.2. With these new vertices we may now repeat our previous procedure.
Thus in the end we arrive at graphs with one vertex only and no tadpoles. But
the unitarity of the associated on-shell S-matrices was established in Section
2, so this property in the general case follows from Theorem 4.1. This is our
third proof of unitarity. As an illustration we consider the following

\begin{figure}[ht]
\centerline{
\unitlength1mm
\begin{picture}(90,60)
\put(40,30){\circle*{2}}
\put(80,30){\circle{2}}
\put(10,30){\line(1,0){30}}
\put(60,30){\oval(40,40)}
\put(30,30){\vector(-1,0){10}}
\put(25,31){1}
\put(60,50){\vector(1,0){5}}
\put(60,51){2}
\end{picture}}
\caption{\label{fig:tadpole}The graph $\Gamma$ with $n=1$ and $m=1$:
a closed loop (i.e. tadpole) of length $a$ plus one external line. The open circle denotes
the vertex added.}
\end{figure}
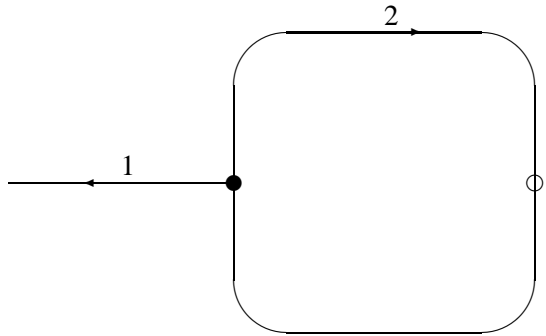
\begin{ex} Consider the Laplacian on the graph $\Gamma$ depicted in Fig. \ref{fig:tadpole}
 with a circle of length $a$ and the same boundary conditions
at the 3-vertex as in Example 2.4. Then
\begin{equation*}
S_\mathrm{tadpole}(E)=e^{i\sqrt{E}a}\frac{e^{-i\sqrt{E}a}-3}{e^{i\sqrt{E}a}-3}.
\end{equation*}
An easy calculation shows that
\begin{equation*}
S_\mathrm{tadpole}(E)=S_{2}^\mathrm{free}(E)*_{p=2}V(a)S(E)V(a),
\end{equation*}
where $S(E)$ is the on-shell S-matrix of Example 2.4 and
\begin{equation*}
V(a)=\left( \begin{array}{ccc}e^{i\sqrt{E}a/2}&0&0\\
                              0&e^{i\sqrt{E}a/2}&0\\
                              0&0&1 \end{array} \right).
\end{equation*}
In this case $K_{1}$ equals
\begin{equation*}
K_{1}=(1-\frac{1}{3}e^{i\sqrt{E}a})^{-1}(1-e^{i\sqrt{E}a})^{-1}
 \left( \begin{array}{cc} 1-\frac{2}{3}e^{i\sqrt{E}a}&
        -\frac{1}{3}e^{i\sqrt{E}a}\\
     -\frac{1}{3}e^{i\sqrt{E}a}&1-\frac{2}{3}e^{i\sqrt{E}a} \end{array} \right),
\end{equation*}
which is singular for $E=\frac{(2\pi n)^{2}}{a^{2}}$.
\end{ex}

The feature exhibited in this example (see also Example 3.2) may be generalized as
follows. Let $\Gamma$ be a graph with local boundary conditions at each vertex which are
scale invariant in the sense of Corollary 2.3. Then by \eqref{compos} the $E$ dependence
of the on-shell S-matrix $S_{\Gamma}(E)$ for this graph only enters through the lengths
$\underline{a}$ in the form $\exp(i\sqrt{E}a_{j})$. In particular
$\lim_{\underline{a}\rightarrow\,0} S_{\Gamma}(E)$ is independent of the energy. This
observation might be helpful in deciding which boundary conditions might be physically
realized in a given experimental context.

\section{Conclusions and Outlook}

In this article we have established unitaritity of the S-matrix for arbitrary finite
quantum wires with a Hamiltonian given by an arbitrary self-adjoint extension of the
Laplace operator. The explicit determination of the on-shell S-matrix has been reduced to
a finite matrix problem, which thus is accessible to computer calculations. A quantum wire
with two open ends but arbitrary interior  and arbitrary boundary conditions may be viewed
as a theory with point interaction and an internal structure (see e.g.
\cite{Pavlov1,Pavlov2}). Ultimately relativistic, local quantum field theories provide the
appropriate set-up for considering point-like interactions and internal structures. Thus
for example the $\Phi^{4}$-theory is the quantum field theoretic version of the
$\delta$-potential. The composition rule established in Section 4 gives a generalization
to arbitrary graphs of the Aktosun factorization formula \cite{Aktosun} for potential
scattering on the line.

Our approach offers several generalizations. First it is possible to
 introduce a potential $V$ on the entire wire thus replacing
$-\Delta(A,B)$ by $-\Delta(A,B)+V$ (see e.g. \cite{GPS}, \cite{Exner96b}).
Although solutions for the on-shell S-matrix in closed form like above may not
be obtainable in general, most structural properties should still hold,
provided the potential is sufficiently strongly decaying at the infinities.
Also one may construct lattice models in the following way. To each lattice
site $j\in\Z^{k}$ in $\R^{k}$ one may associate an S-matrix with $2k$ open ends
and connect neighboring lattice sites accordingly (see e.g.
\cite{Avishai:Luck,Gratus,Exner95}). This would lead to the possibility of
considering infinite quantum wires - obtained as `thermodynamic' limits of
finite quantum wires - and study their conductance properties in the spirit of
e.g. the analysis in \cite{Avishai:Luck}. Thus if one takes the same S-matrices
at each site the resulting theory will be translation invariant. The
generalized star product also offers the possibility of introducing a transfer
matrix theory (see again \cite{Avishai:Luck}). For the case $k=1$ this will
generalize the Kronig-Penney model \cite{Avron:Exner:Last,KS2}. Finally the
S-matrices at the lattice sites could vary stochastically allowing the study of
percolation effects.

The present discussion of Laplace operators on graphs could also be used to
study Brownian motion and the associated diffusion process. More precisely we
expect the heat kernel $\exp(t\Delta(A,B))(x,y)$ to be nonnegative if the
boundary conditions $(A,B)$ are real. It would be interesting to obtain a
representation in the spirit of the Selberg and Gutzwiller trace formula. If
the boundary conditions are not real, reflecting the presence of a magnetic
field say, we expect a corresponding Ito formula to be valid and Aharonov-Bohm
like effects to show up.

\section*{Acknowledgements}

The authors are indebted to Y. Avron, Th. Br\"ocker, J.M. Combes, M. Karowski,
P. Kurasov, and L. Sadun for valuable comments. We thank G. Dell'Antonio for
pointing out Ref. \cite{Novikov} and to S.P. Novikov for sending us the
manuscript \cite{Novikov} prior to its publication. Suggestions of anonymous
referees are also acknowledged.

\setcounter{equation}{0}
\section*{Appendix A}
\renewcommand{\theequation}{A\arabic{equation}}
\setcounter{equation}{0}
In this Appendix we will describe $\Delta(A,B)$ defined on a graph with a single vertex
from the viewpoint of von Neumann's extension theory (see e.g. \cite{RS}). According
to von Neumann's theorem any self-adjoint extension $\Delta(A,B)$ of $\Delta^0$ can be
uniquely parametrized by a linear isometric isomorphism $\cW_{A,B}:\
\Ker(-{\Delta^0}^\dagger-i)\rightarrow
\Ker(-{\Delta^0}^\dagger+i)$ according to the formula
\begin{eqnarray*}
&&\cD(\Delta(A,B)) = \left\{\psi+\psi_++\cW_{A,B}\psi_+|\ \psi\in\cD(\Delta^0),
\psi_+\in\Ker(-{\Delta^0}^\dagger-i) \right\},\\
&&-\Delta(A,B)(\psi+\psi_++\cW_{A,B}\psi_+)=-\Delta^0\psi+i\psi_+-i\cW_{A,B}\psi_+.
\end{eqnarray*}
To express
$\cW_{A,B}$ in terms of the matrices $A$ and $B$ we compare $\Delta(A,B)$ with another
self-adjoint extension $\Delta(A',B')$ of $\Delta^0$. For definiteness we take $A'=\1$ and
$B'=0$, which corresponds to the Dirichlet boundary conditions (but another choice would also
be possible).

Let $u_j\in\Ker(-{\Delta^0}^\dagger-i)$ and $v_j\in\Ker(-{\Delta^0}^\dagger+i)$,
$j=1,\ldots,n$ be given as
\begin{equation}\label{uv}
(u_j(x))_k=\delta_{jk}2^{1/4}e^{\frac{1}{\sqrt{2}}(-1+i)x},\quad
(v_j(x))_k=\delta_{jk}2^{1/4}e^{\frac{1}{\sqrt{2}}(-1-i)x}, \quad k=1,\ldots,n.
\end{equation}
One can easily verify that $\{u\}_{j=1}^n$ and $\{v\}_{j=1}^n$ are orthonormal bases for
$\Ker(-{\Delta^0}^\dagger-i)$ and $\Ker(-{\Delta^0}^\dagger+i)$, respectively. Denote by
$W_{A,B}$ the unitary matrix representation of $\cW_{A,B}$ with respect to the bases
$\{u\}_{j=1}^n$ and $\{v\}_{j=1}^n$, i.e.
\begin{displaymath}
\cW_{A,B}u_j=\sum_{k=1}^n\left(W_{A,B}\right)_{kj}v_k.
\end{displaymath}
It is easy to see that $W_{A'=\1,B'=0}=-\1$. Our strategy will be to apply
Krein's formula (see e.g. \cite{Achieser:Glasmann}) to obtain $W_{A,B}$ in terms of the
matrices $A$ and $B$. To construct $\Delta(A,B)$ in the sense of
von Neumann we proceed in two steps.
In the first step let $\widetilde{\Delta}$ be the maximal common part of $\Delta(A,B)$ and
$\Delta(A'=\1,B'=0)$, i.e. $\widetilde{\Delta}$ is the largest closed extension of
$\Delta^0$ with
\begin{displaymath}
\cD(\widetilde{\Delta})=\cD(\Delta(A,B))\cap \cD(\Delta(A'=\1,B'=0))=
\left\{\psi|\ \psi(0)=0, B\psi'(0)=0 \right\}.
\end{displaymath}
Obviously, $\widetilde{\Delta}$ is a symmetric operator with defect indices $(r,r)$,
$r=\Rank B$ $(0\leq r\leq n)$. Also one has $\Ker(-{\widetilde{\Delta}}^\dagger\mp
i)\subseteq\Ker(-{\Delta^0}^\dagger\mp i)$ such that we may decompose
\begin{equation}\label{dec}
\Ker(-{\Delta^0}^\dagger\mp i)=\Ker(-{\widetilde{\Delta}}^\dagger\mp i)\oplus\cM_\pm.
\end{equation}
Performing, if necessary, a renumeration of the basis elements we may assume
 that $\{u\}_{j=1}^r$
and $\{v\}_{j=1}^r$ are bases for $\Ker(-{\widetilde{\Delta}}^\dagger-i)$ and
$\Ker(-{\widetilde{\Delta}}^\dagger+i)$ respectively.

Von Neumann's theorem describes $\widetilde{\Delta}$ as a symmetric extension of
$\Delta^0$. All such extensions are parametrized by linear partial isometries $\cW':\
\Ker(-{\Delta^0}^\dagger-i)\rightarrow\Ker(-{\Delta^0}^\dagger+i)$. For the case at hand it
is easy to see that the isometry
 corresponding to $\widetilde{\Delta}$ defines an isometric isomorphism $\widetilde{\cW}$:
 $\cM_+\rightarrow\cM_-$ and therefore
\begin{eqnarray*}
&&\cD(\widetilde{\Delta})=\left\{\psi+\psi_++\widetilde{\cW}\psi_+|\ \psi\in\cD(\Delta^0),
\psi_+\in\cM_+\right\},\\
&&-\widetilde{\Delta}(\psi+\psi_++\widetilde{\cW}\psi_+)=
-\Delta^0\psi+i\psi_+-i\widetilde{\cW}\psi_+.
\end{eqnarray*}
Denoting by $\widetilde{W}$ the matrix
representation of $\widetilde{\cW}$ in the bases $\{u\}_{j=r+1}^n$ and $\{v\}_{j=r+1}^n$
we get that
$\widetilde{W}=-\1$.

In the second step let $\widetilde{\cW}_{A,B}:\
\Ker(-{\widetilde{\Delta}}^\dagger-i)\rightarrow\Ker(-{\widetilde{\Delta}}^\dagger+i)$ be
the linear isometric isomorphism parametrizing $\Delta(A,B)$ as a self-adjoint extension of
$\widetilde{\Delta}$. Denote by $\widetilde{W}_{A,B}$ its unitary matrix representation
with respect to the bases $\{u\}_{j=1}^r$ and $\{v\}_{j=1}^r$. Again it is easy to see
that $\widetilde{W}_{A'=\1,B'=0}=-\1$ ($r\times r$ matrix). Also from the discussion above
it follows that
\begin{equation}\label{decW}
W_{A,B}=\widetilde{W}_{A,B}\oplus(-\1)
\end{equation}
with respect to the decomposition (\ref{dec}).

Krein's formula (see e.g. \cite{Achieser:Glasmann}) now states that the difference of the
resolvents of $-\Delta(A,B)$ and $-\Delta(A'=\1,B'=0)$ at the point $z=i$ is given by
\begin{equation}\label{Krein}
R_{A,B}(i)-R_{A'=\1,B'=0}(i)=\sum_{j,k=1}^r \widetilde{P}_{jk}(v_k,\cdot)v_j,
\end{equation}
with some $r\times r$ matrix $\widetilde{P}$ of maximal rank. By Corollary B.3 of
\cite{Gesztesy} it follows that
\begin{equation}\label{Gesztesy}
\widetilde{P}=\frac{i}{2}(\1+\widetilde{W}_{A,B}^{-1}).
\end{equation}
Obviously, relation (\ref{Krein}) can be rewritten in the form
\begin{equation}\label{Krein1}
R_{A,B}(i)-R_{A'=\1,B'=0}(i)=\sum_{j,k=1}^n P_{jk}(v_k,\cdot)v_j,
\end{equation}
with $P$ being $n\times n$ matrix of rank $r$ such that $P=\widetilde{P}\oplus 0$
with respect to the decomposition (\ref{dec}). Thus from (\ref{Gesztesy}) and (\ref{decW}) it
follows that
\begin{equation}\label{Gesztesy1}
P=\frac{i}{2}(\1+W_{A,B}^{-1}).
\end{equation}

The resolvent $R_{A'=\1,B'=0}(i)$ of $\Delta(A'=\1,B'=0)$ can be given explicitly. Its
integral kernel (Green's function) for $x<y$ has the form
\begin{displaymath}
R_{A'=\1,B'=0}(x,y;i)
=\1\frac{\sin\sqrt{i}x}{\sqrt{i}}e^{\frac{1}{\sqrt{2}}(-1+i)y},
\end{displaymath}
such that
\begin{displaymath}
R_{A'=\1,B'=0}(0,y;i)=0,\qquad \frac{\partial R_{A'=\1,B'=0}}{\partial x}(0,y;i)=\1
e^{\frac{1}{\sqrt{2}}(-1+i)y}.
\end{displaymath}
Since the boundary conditions take the form
\begin{displaymath}
AR_{A,B}(0,y;i)+B\frac{\partial R_{A,B}}{\partial x}(0,y;i)=0
\end{displaymath}
for all $y>0$ in terms of the resolvents, from (\ref{Krein1}) we obtain
\begin{displaymath}
\left(A+\frac{1}{\sqrt{2}}(-1+i)B\right)P=-\frac{1}{\sqrt{2}}B.
\end{displaymath}
Comparing this with (\ref{Gesztesy1}) we obtain
\begin{displaymath}
W_{A,B}^{-1}=-\left(A+\frac{1}{\sqrt{2}}(-1+i)B\right)^{-1}
\left(A-\frac{1}{\sqrt{2}}(1+i)B\right).
\end{displaymath}
Note that $(\widehat{A},\widehat{B})$ given as
\begin{equation*}
\widehat{A}=A-\frac{1}{\sqrt{2}}B,\;\widehat{B}= \frac{1}{\sqrt{2}}B
\end{equation*}
also defines a maximal isotropic subspace and
$W_{A,B}^{-1}=S_{\widehat{A},\widehat{B}}(E=1)$ (the S-matrix for the single vertex
theory) holds, such that in particular $W_{A,B}$ is unitary and satisfies
$W_{A,B}=W_{CA,CB}$ for any invertible $C$, as it should be. As an example, for $A=0$ and
$B=\1$ (Neumann boundary conditions) this gives $W_{A=0,B=\1}=i\1$, again as it should be.

\setcounter{equation}{0}
\section*{Appendix B}
\renewcommand{\theequation}{B\arabic{equation}}
\setcounter{equation}{0}

Here we give an alternative `analytic' proof of Theorem 3.3, i.e. of the
unitarity of the on-shell S-matrix. We resort to an argument which is a
modification of well known arguments used to prove orthogonality relations for
improper eigenfunctions of Schr\"{o}dinger Hamiltonians (see also the remark at the
end of this appendix) and which is instructive in its own right. For this
purpose we introduce Hilbert spaces $\cH_{R}$ indexed by $R>0$ and given as
\begin{equation*}
    \cH_{R}=\oplus_{e\in\cE}\cH_{e,R}\oplus_{i\in\cI}\cH_{i},
\end{equation*}
where the spaces $\cH_{i}$ are as before and where $\cH_{e,R}=L^{2}([0,R])$ for all
$e\in\cE$. Then $\psi^{k}(\cdot,E)\in\cH_{R}$ for all $e\in\cE$ and all $0<R<\infty$. Let
$\langle\;,\;\rangle_{R}$ denote the canonical scalar product on $\cH_{R}$. Also we have
$\Delta\psi^{k}(\cdot,E)=-E\psi^{k}(\cdot,E)$ such that for all $E,E^{\prime}
\in\R\setminus\Sigma_{A,B}$ and all $k,l\in\cE$
\begin{eqnarray}\label{556}
\langle\Delta\psi^{k}(\cdot,E),\psi^{l}(\cdot,E^{\prime})\rangle_{R}-
\langle\psi^{k}(\cdot,E),\Delta\psi^{l}(\cdot,E^{\prime})\rangle_{R}\nonumber\\
 =(E^{\prime}-E)\langle\psi^{k}(\cdot,E),\psi^{l}(\cdot,E^{\prime})\rangle_{R}.
\end{eqnarray}
Now all terms in this equation are smooth in $E,E^{\prime}\in\R\setminus\Sigma_{A,B}$ and
$R$. We may therefore take the limit $E^{\prime}\rightarrow E$. This gives for the right
hand side of \eqref{556}
\begin{equation*}
\lim_{E^{\prime}\rightarrow E}
(E^{\prime}-E)\langle\psi^{k}(\cdot,E),\psi^{l}(\cdot,E^{\prime})\rangle_{R}=0.
\end{equation*}
On the other hand we may evaluate the left hand side of \eqref{556}
by performing a partial integration. Since both $\psi^{k}(\cdot,E)$ and
$\psi^{l}(\cdot,E^{\prime})$ satisfy the boundary conditions $(A,B)$, the only
 contributions arise at the $n$ `new' boundaries $x=R$, i.e.
\begin{eqnarray*}
\begin{array}{cc}
{}&\langle\Delta\psi^{k}(\cdot,E),\psi^{l}(\cdot,E^{\prime})\rangle_{R}-
\langle\psi^{k}(\cdot,E),\Delta\psi^{l}(\cdot,E^{\prime})\rangle_{R}\\
\\
=&\displaystyle\sum_{e\in\cE}\left( \bar{\psi}_{e}^{k^{\;\prime}}(R,E)\;
 \psi^{l}_{e}(R,E^{\prime})
-\bar{\psi}^{k}_{e}(R,E)\;\psi^{l^{\;\prime}}_{e}(R,E^{\prime})
    \right)
\end{array}\\
\begin{array}{cc}=& -i(\sqrt{E^{\prime}}+\sqrt{E})
e^{i(\sqrt{E^{\prime}}-\sqrt{E})R}\;
 \displaystyle\sum_{e\in\cE}
 \overline{S_{ek}(E)}S_{el}(E^{\prime})\\
\\
{}&+i(\sqrt{E^{\prime}}+\sqrt{E})
  e^{-i(\sqrt{E^{\prime}}-\sqrt{E})R}\;\delta_{kl}\\
\\
{}&+i(\sqrt{E^{\prime}}-\sqrt{E})e^{i(\sqrt{E^{\prime}}+\sqrt{E})R}\;
  \overline{S_{lk}(E)}\\
\\
{}& -i(\sqrt{E^{\prime}}-\sqrt{E})
  e^{i(\sqrt{E^{\prime}}+\sqrt{E})R}\;
  S_{kl}(E^{\prime}). \end{array}
\end{eqnarray*}
The limit of the right hand side as
$E^{\prime}\rightarrow E$ equals
\begin{equation*}
 -2i\sqrt{E}\left(\sum_{e\in\cE}\overline{S_{ek}(E)}S_{el}(E)-\delta_{kl}\right),
\end{equation*}
which by the preceding arguments is zero, concluding this second proof of unitarity.

We could have chosen $E^{\prime}=E$ from the very beginning. However, the present
discussion may be applied to show general orthogonality properties for the improper
eigenfunctions $\psi^{k}(\cdot,E)$. This is achieved by first dividing
\eqref{556} by $E^{\prime}-E$ and then taking the limit $R\rightarrow\infty$ in the sense
of distributions in $E^{\prime}$ and $E$.

\section*{Appendix C}
\renewcommand{\theequation}{B\arabic{equation}}
\setcounter{equation}{0}

This appendix is devoted to a proof of Theorem 4.1. We start with the following
observation. For $U^{\prime}$ written in block form as in \eqref{54} we define
a map $U^{\prime}\rightarrow U^{\prime^{\tau}}$ with
$\tau=\tau(n^{\prime}-p,p)$ given as
\begin{equation*}
          U^{\prime^{\tau}}=\left(\begin{array}{cc}
           U^{\prime}_{22}& U^{\prime}_{21}\\
           U^{\prime}_{12}&U^{\prime}_{11}\end{array}\right),
\end{equation*}
which amounts to interchanging the first $n^{\prime}-p$ indices with the last
$p$ indices while keeping the order of the indices otherwise fixed.
$U^{\prime\prime^{\tau}}$ and $U^{\tau}$ are defined analogously with
$\tau=\tau(p,n^{\prime\prime}-p)$ and
$\tau=\tau(n^{\prime}-p,n^{\prime\prime}-p)$ respectively. From the definition
of $*_{V}$ it follows immediately that the following `transposition law' holds
\begin{equation*}
U^{\tau}=U^{\prime\prime^{\tau}}*_{V^{-1}}U^{\prime^{\tau}}
\end{equation*}
whenever $U=U^{\prime}*_{V}U^{\prime\prime}$. To prove Theorem 4.1 we have to show
that the following four relations hold
\begin{eqnarray*}
     U^{\dagger}_{11}\,U_{11}+U^{\dagger}_{21}\,U_{21}&=&\1,\\
     U^{\dagger}_{11}\,U_{12}+U^{\dagger}_{21}\,U_{22}&=&0,\\
     U^{\dagger}_{12}\,U_{12}+U^{\dagger}_{22}\,U_{22}&=&\1,\\
     U^{\dagger}_{12}\,U_{11}+U^{\dagger}_{22}\,U_{21}&=&0.
\end{eqnarray*}
By our previous observation it suffices to prove only the first two relations. To prove
the first one, we insert the definition of $U$ and obtain
\begin{eqnarray*}
U^{\dagger}_{11}\,U_{11}+U^{\dagger}_{21}\,U_{21}
&=&U^{\prime\dagger}_{11}\,U^{\prime}_{11}
 +U^{\prime\dagger}_{11}\,U^{\prime}_{12}\,
 K_{2}\,U^{\prime\prime}_{11}\,V\,U^{\prime}_{21}\\
 &&+U^{\prime\dagger}_{21}\,V^{-1}\,U^{\prime\prime\dagger}_{11}\,
 K_{2}\,U^{\prime\dagger}_{12}\,U^{\prime}_{11}\\
&&+U^{\prime\dagger}_{21}\,V^{-1}U^{\prime\prime\dagger}_{11}\,
K^{\dagger}_{2}\,U^{\prime\dagger}_{12}\,U^{\prime}_{12}\,
K_{2}\,U^{\prime\prime}_{11}\,V\,U^{\prime}_{21}\\
&&+U^{\prime\dagger}_{21}\,K^{\dagger}_{1}\,
U^{\prime\prime\dagger}_{21}\,U^{\prime\prime}_{21}\,K_{1}\,U^{\prime}_{21}\\
&=&\sum^{5}_{i=1}a_{i}.
\end{eqnarray*}
We now use the unitarity relations for $U^{\prime}$ and $U^{\prime\prime}$.
This gives
\begin{eqnarray*}
\sum^{5}_{i=2}a_{i}&=&U^{\prime\dagger}_{21}\{
   -U^{\prime}_{22}\, K_{2}\, U^{\prime\prime}_{11}\, V
   -V^{-1}\, U^{\prime\prime\dagger}_{11}\,K^{\dagger}_{2}\, U^{\prime}_{22}\\
 &&+V^{-1}\,U^{\prime\prime\dagger}_{11}\,K^{\dagger}_{2}\,U^{\prime\dagger}_{12}\,
  U^{\prime}_{12}\,K_{2}\,U^{\prime\prime}_{11}\,V+
K^{\dagger}_{1}\,U^{\prime\prime\dagger}_{21}\,U^{\prime\prime}_{21}\,K_{1}\}
   U^{\prime}_{21}\\
 &=&U^{\prime\dagger}_{21}\{ -U^{\prime}_{22}\,
  K_{2}\,U^{\prime\prime}_{11}\,V-
V^{-1}\,U^{\prime\prime\dagger}_{11}\,K^{\dagger}_{2}\,U^{\prime}_{22}
+V^{-1}\,U^{\prime\prime\dagger}_{11}\,K^{\dagger}_{2}\,K_{2}\,U^{\prime\prime}_{11}\,V\\
&&-V^{-1}\,U^{\prime\prime\dagger}_{11}\,K^{\dagger}_{2}\,U^{\prime\dagger}_{22}\,
U^{\prime}_{22}\, K_{2}\, U^{\prime\prime}_{11}\,V+
K^{\dagger}_{1}\,K_{1}-K^{\dagger}_{1}\,U^{\prime\prime\dagger}_{11}\,
U^{\prime\prime}_{11}\,K_{1}\} U^{\prime}_{21}.
\end{eqnarray*}
To establish $\sum_{i=1}^{5}a_{i}=\1$ it therefore suffices to show that the expression in
braces equals $\1$. Using one of the relations in \eqref{345} and its adjoint we have
\begin{eqnarray*}
 K^{\dagger}_{1}K_{1}
&=&\1+U^{\prime}_{22}\,K_{2}\,U^{\prime\prime}_{11}\,V
+V^{-1}\,U^{\prime\prime\dagger}_{11}\, K^{\dagger}_{2}\,U^{\prime\dagger}_{22}\\
&&+V^{-1}\,U^{\prime\prime\dagger}_{11}\,K^{\dagger}_{2}\,
U^{\prime\dagger}_{22}\,U^{\prime}_{22}\,K_{2}\,U^{\prime\prime}_{11}\,V
\end{eqnarray*}
and
\begin{eqnarray*}
K^{\dagger}_{1}\,U^{\prime\prime\dagger}_{11}\,U^{\prime\prime}_{11}\,K_{1}&=&
V^{-1}\,(\1+U^{\prime\prime\dagger}_{11}\,K^{\dagger}_{2}\,U^{\prime\dagger}_{22}\,
V^{-1})\\ &&U^{\prime\prime\dagger}_{11}\,U^{\prime\prime}_{11}
(\1+V\,U^{\prime}_{22}\,K_{2}\,U^{\prime\prime}_{11})V.
\end{eqnarray*}
Hence it suffices to show that
\begin{equation*}
-(\1+U^{\prime\prime\dagger}_{11}\,K^{\dagger}_{2}\,U^{\prime\dagger}_{22}\,
  V^{-1})U^{\prime\prime\dagger}_{11}\,U^{\prime\prime}_{11}
 (\1+VU^{\prime}_{22}\,K_{2}\,U^{\prime\prime}_{11})
+U^{\prime\prime\dagger}_{11}\,K^{\dagger}_{2}\, K_{2}\,U^{\prime\prime}_{11}=0.
\end{equation*}
To show this it suffices in turn to prove that
\begin{eqnarray*}
 K^{\dagger}_{2}K_{2}
&=&\1+K^{\dagger}_{2}\,U^{\prime\dagger}_{22}\,V^{-1}\,U^{\prime\prime\dagger}_{11}
  +U^{\prime\prime}_{11}\,V\,U^{\prime}_{22}\,K_{2}\\
 &&+K^{\dagger}_{2}\,U^{\prime\dagger}_{22}\,V^{-1}\,U^{\prime\prime\dagger}_{11}\,
       U^{\prime\prime}_{11}\,V\,U^{\prime}_{22}\,K_{2}.
\end{eqnarray*}
But this relation follows by inserting the relation
\begin{equation*}
K_{2}=V^{-1}+V^{-1}\,U^{\prime\prime}_{11}\,V\,U^{\prime}_{22}\,K_{2}
\end{equation*}
and its adjoint into the left hand side. To sum up we have proved the first
 of the unitarity relations for $U$. To prove the second relation we again
insert
the definition of $U$, use the unitarity relations for $U^{\prime}$ and
$U^{\prime\prime}$ twice and obtain
\begin{eqnarray*}
  U^{\dagger}_{11}\,U_{12}+U^{\dagger}_{21}\,U_{22}&=&
U^{\prime\dagger}_{21}\{-U^{\prime}_{22}\,K_{2}
 +V^{-1}\,U^{\prime\prime\dagger}_{11}\,K^{\dagger}_{2}\,K_{2}\\
&&-V^{-1}\,U^{\prime\prime\dagger}_{11}\,K^{\dagger}_{2}\,
 U^{\prime\dagger}_{22}\,
 U^{\prime}_{22}\,K_{2}\\
 &&-K^{\dagger}_{1}\,U^{\prime\prime\dagger}_{11}
  +K^{\dagger}_{1}\,K_{1}\,U^{\prime}_{22}\,V^{-1}\\
 &&  -K^{\dagger}_{1}\,U^{\prime\prime\dagger}_{11}\,U^{\prime\prime}_{11}\,
  K_{1}\,U^{\prime}_{22}\,V^{-1}\} U^{\prime\prime}_{12}.
\end{eqnarray*}
Hence it suffices to show that the expression in braces $(=a_{6})$ vanishes.
But
\begin{eqnarray*}
-K^{\dagger}_{1}\,U^{\prime\prime\dagger}_{11}-K^{\dagger}_{1}\,
U^{\prime\prime\dagger}_{11}\,U^{\prime\prime}_{11}\,
  K_{1}\,U^{\prime}_{22}\,V^{-1}&=&-K^{\dagger}_{1}\,
  U^{\prime\prime\dagger}_{11}\,V\,K_{2}\\
 &=&-V^{-1}(\1+U^{\prime\prime\dagger}_{11}\,
K^{\dagger}_{2}\,U^{\prime\dagger}_{22}\,V^{-1}) U^{\prime\prime\dagger}_{11}\,V\,K_{2}\\
&=&-V^{-1}\,U^{\prime\prime\dagger}_{11}\,V\,K_{2}
-V^{-1}\,U^{\prime\prime\dagger}_{11}(K^{\dagger}_{2}-V)K_{2}\\
 &=&-V^{-1}\,U^{\prime\prime\dagger}_{11}\,K^{\dagger}_{2}\,K_{2}
\end{eqnarray*}
implies
\begin{equation*}
a_{6}=-U^{\prime}_{22}\,K_{2}-V^{-1}\,U^{\prime\prime\dagger}_{11}\,
      K^{\dagger}_{2}\,U^{\prime\dagger}_{22}\,U^{\prime}_{22}\,K_{2}
     +K^{\dagger}_{1}\,K_{1}\,U^{\prime}_{22}\,V^{-1}.
\end{equation*}
The chain of equalities
\begin{eqnarray*}
K^{\dagger}_{1}\,K_{1}\,U^{\prime}_{22}\,V^{-1}&=&
(\1+V^{-1}\,U^{\prime\prime\dagger}_{11}\,K^{\dagger}_{2}\,U^{\prime\dagger}_{22})
(\1+U^{\prime}_{22}\,K_{2}\,
 U^{\prime\prime}_{11}\,V) U^{\prime}_{22}\,V^{-1}\\
&=&U^{\prime}_{22}\,V^{-1}+U^{\prime}_{22}\,K_{2}\,U^{\prime\prime}_{11}\,V\,
U^{\prime}_{22}\,V^{-1}\\ &&+V^{-1}\,U^{\prime\prime\dagger}_{11}\,
 K^{\dagger}_{2}\,U^{\prime\dagger}_{22}\,U^{\prime}_{22}\,
 K_{2}\,U^{\prime\prime}_{11}\, V\, U^{\prime}_{22}\,V^{-1}\\
 &&+V^{-1}\, U^{\prime\prime\dagger}_{11}\, K^{\dagger}_{2}\,
U^{\prime\dagger}_{22}\, U^{\prime}_{22}\, V^{-1}\\ &=&U^{\prime}_{22}\,
V^{-1}+U^{\prime}_{22} (K_{2}-V^{-1})\\  &&+V^{-1}\, U^{\prime\prime\dagger}_{11}
K^{\dagger}_{2}\, U^{\prime\dagger}_{22}\, U^{\prime}_{22}(K_{2}-V^{-1})\\ &&+V^{-1}\,
U^{\prime\prime\dagger}_{11}\, K^{\dagger}_{2}\, U^{\prime\dagger}_{22}\,
U^{\prime}_{22}\, V^{-1}\\  &=&U^{\prime}_{22}\, K_{2}+V^{-1}\,
U^{\prime\prime\dagger}_{11}\, K^{\dagger}_{2}\, U^{\prime\dagger}_{22}\,
U^{\prime}_{22}\, K_{2}
\end{eqnarray*}
finally leads to $a_{6}=0$ as desired concluding the proof of Theorem 4.1.

\markright{References}

\end{document}